\documentclass{aa}  

\usepackage{graphicx}
\usepackage{txfonts}

\usepackage{natbib}
\bibpunct{(}{)}{;}{a}{}{,} 

\begin{document}

   \title{The Galactic WC and WO stars}

   \subtitle{The impact of revised distances from Gaia DR2 and their role as massive black hole progenitors}

   \author{A.A.C. Sander\inst{\ref{inst1},\ref{inst2}}
          \and
          W.-R. Hamann\inst{\ref{inst1}}
          \and
					H. Todt\inst{\ref{inst1}}
          \and
          R. Hainich\inst{\ref{inst1}}
   			  \and
   			  T. Shenar\inst{\ref{inst1},\ref{inst3}}
					\and
					V. Ramachandran\inst{\ref{inst1}}
					\and
   			  L.M. Oskinova\inst{\ref{inst1}}
          }

   \institute{Institut f\"ur Physik und Astronomie, Universit\"at Potsdam,
              Karl-Liebknecht-Str. 24/25, D-14476 Potsdam, Germany\\
              \email{ansander@astro.physik.uni-potsdam.de}\label{inst1}
							\and
							Armagh Observatory and Planetarium, College Hill, Armagh,
              BT61 9DG, Northern Ireland\label{inst2}
							\and
							Institute of Astrophysics, KU Leuven, Celestijnenlaan 200 D, 3001, Leuven, Belgium\label{inst3}
             }

  \date{Received 25 June 2018 / Accepted 26 November 2018}

\abstract{
  Wolf-Rayet stars of the carbon sequence (WC stars) are an important cornerstone in
	the late evolution of massive stars before their core collapse. As core-helium burning, hydrogen-free 
	objects with huge mass-loss, they are likely the last observable stage before collapse and thus promising 
	progenitor candidates for type Ib/c supernovae. Their strong mass-loss furthermore provides challenges and 
	constraints to the theory of radiatively driven winds. 
	Thus, the determination of the WC star parameters is of major importance for several astrophysical fields.
	With Gaia DR2, for the first time parallaxes for a large sample of Galactic WC stars are available, 
	removing major uncertainties inherent to earlier studies. 
	
	In this work, we re-examine a previously studied sample of WC stars
	to derive key properties of the Galactic WC population. All quantities 
	depending on the distance are updated, while the underlying spectral analyzes remain untouched.
	Contrasting earlier assumptions, our study yields that WC stars of the same subtype can significantly vary in absolute magnitude.
	
	With Gaia DR2, the picture of the Galactic WC population becomes more complex: We obtain luminosities 
	ranging from $\log L/L_\odot = 4.9$ up to $6.0$ with one outlier (WR 119) having $\log L/L_\odot = 4.7$. 
	This indicates that the WC stars are likely formed from a broader
	initial mass range than previously assumed. We obtain mass-loss rates ranging 
	between $\log \dot{M} = -5.1$ and $-4.1$, with $\dot{M} \propto L^{0.68}$
	and a linear scaling of the modified wind momentum with luminosity.  
	
	We discuss the implications for stellar evolution, including unsolved issues regarding the need
	of envelope inflation to address the WR radius problem, and the open questions in regard to the connection
	of WR stars with Gamma-ray bursts. WC and WO stars are progenitors of massive black holes, collapsing either
	silently or in a supernova that most-likely has to be preceded by a WO stage.
}



\maketitle


\section{Introduction}
  \label{sec:intro}
	
	Classical Wolf-Rayet (WR) stars are descendants of evolved, massive stars that have
	lost their outermost layers. 
	Officially defined in a purely spectroscopic way by the presence of certain emission 
	lines in their optical spectrum \citep[e.g.][]{HS1966,S1968WRclass,Crowther+1998}, the
  term ``classical'' was added more recently to distinguish them from other
	objects with similar emission-line phenomena, e.g. central stars of planetary nebulae
	or the so-called O-stars on steroids, in other words very massive early-type stars with strong
	winds that are still in the stage of core-hydrogen burning.
	Based on the presence of prominent emission lines of nitrogen or carbon, WR stars are 
	sorted into the nitrogen (WN), carbon (WC), and oxygen (WO) sequence with subclasses defined by the ratios of characteristic
	emission lines. While some WN stars have considerable amounts of hydrogen, all known WC and 
	WO stars do not reveal any measurable amount of hydrogen in their spectra. 
	
	The WO subtype is defined by the (strong) presence of \ion{O}{vi}\,$\lambda$3811\,\AA\ (compared to \ion{O}{v}\,$\lambda$5590\,\AA)
	that is usually seen either as an extension of the WC sequence to hotter
	temperatures or a distinct spectroscopic sequence of more evolved stars. Current evolutionary calculations
	\citep[e.g.][]{Groh+2014} and empirical studies \citep[e.g.][]{Sander+2012,Tramper+2015} 
	support a picture that combines both approaches: Late WO subtypes have 
	chemical compositions quite similar to early WC subtypes while early WO subtypes have higher
	oxygen abundances and indeed reflect the last stage before core collapse that is observable for a perceptible duration of time.
	
	Wolf-Rayet stars have an enormous influence on their environment. Due to their high temperatures, they
	produce large numbers of ionizing photons. Their huge mass-loss enriches the surrounding
	interstellar medium with metals and injects a significant amount of mechanical energy into it. Few
	WR stars can equal or even outweigh the influence of a whole population of nearby OB stars \citep[e.g.][]{Ramachandran+2018}.
	Thus, quantifying their impact is of particular importance for a variety of astrophysical applications.
	
	As hydrogen-free and thus (at least) core-helium burning objects, the WC and WO stars represent
	a clearly defined stage. In \citet{Sander+2012}, all (putatively) single Galactic WC and WO stars 
	with at that time available spectra were analyzed with the help of the Potsdam Wolf-Rayet (PoWR)
	model atmospheres. To obtain their luminosities, their distances need to be known. While this
	is not much of an issue for WR populations in other galaxies \citep[see, e.g.][]{Hainich+2014,Hainich+2015,Sander+2014,Shenar+2016},
	most of the distances to the WR stars in our Milky Way were not well constrained before Gaia. 
	They are too far away to have a (reliable) HIPPARCOS parallax, and  
	especially the unobscured sources are often relatively isolated. Thus only a small sub-sample 
	in \citet{Sander+2012} had distances that were due to confirmed or suspected cluster or association
	membership. However, even for confirmed cluster members, the distances were not always robust, as also
	the distances to various clusters had been a matter of debate, as for example reflected in the discussion about the Be 87 cluster
	hosting the WO2 star WR 142 \citep[cf. appendix A in][]{Sander+2012}.
	
	Roughly two decades after HIPPARCOS, the Gaia mission \citep{GaiaCollab2016}, now with its second 
	data release \citep{GaiaCollab2018}, provides for the first time parallaxes \citep{Luri+2018} for most of the 
	unobscured Galactic WR stars, covering the full sample of Galactic WC stars from \citet{Sander+2012}. Based on these parallaxes,
	new distances can be derived and we can revise the luminosities and further distance-dependent parameters such as mass-loss
	rates. In this work, we only focus on the impact of the revised distances and
	do not perform any updates of our atmosphere models assigned to each of the
	sample stars. While the PoWR code has undergone considerable extensions during
	the last years \citep[e.g.][]{Shenar+2014,Sander+2015,Sander+2017}, these are
	either not essential for the empirical analysis of WR stars at Galactic metallicity or 
	have only moderate effects on the resulting stellar parameters. Thus, instead of confusing
	the effects of some model details with the impact of the revised distances, we decided to
	focus on the latter. 
	
	In contrast, the PoWR models compiled in the online library\footnote{\texttt{http://www.astro.physik.uni-potsdam.de/PoWR}}
	contain the most recent grids of WC models for Galactic and LMC metallicity, accounting for the
	updates and extensions of the PoWR code in the past years. Besides many minor improvements and additional information
	on the website, the most prominent update compared to the WC grid released with \citet{Sander+2012} is an improved 
	treatment of the superlevels for the iron group elements and the addition of infrared line spectra to the online database. 
	While the deduced stellar parameters for the Galactic WC sample with the newer generation of models are within the
	uncertainty of what we deduced with the older generation, the improved treatment of the iron group elements can have
	a considerable effect on parts of the emergent spectrum for certain parameter regimes, especially in the UV. 
	Thus, for a detailed study of a particular WC star, we recommend to use the newest version of the WC grid.
	
	Based on recent literature, we re-examined the binary status of each object in our sample. 
	WR 14 was reported to exhibit significant variability in the radio domain, which was suggested to originate from 
	binarity \citep{Drissen+1992, deBecker+2013}. However, the spectral analysis compared to other stars of the same
	subtype does not indicate a significant contribution from the companion.
	
	It is generally thought that many if
	not all of the dust-producing WC9 stars are members in (long-period) binaries \citep[e.g.][]{Usov1991, Tuthill+1999}. 
	For WR 65, the new distance leads to an X-ray luminosity of roughly $L_\text{X} \approx 10^{33}\,$erg/s, a value
	typical for colliding wind binaries (CWBs), a suggestion already reported by \citet{OH2008}.
	\citet{Desforges+2017} investigated a large sample of WC9 / WC9d stars for line-flux variability, but could not 
	find evidence for periodicity in their data, with the possible exception of WR 103. However, they cannot exclude 
	the possibility that the stars are members of long-period binaries. 
	The WN/WC transition-type star WR 145 is a WR+O colliding wind binary system which unfortunately was not properly 
	marked in \citet{Sander+2012}. For completeness, we keep it in our sample, but set a mark in Table\,\ref{tab:wcpar}
	and also comment on it in Sect.\,\ref{subsec:notable}. Apart from WR 145, a few objects are suspicious and we discuss
	these in Sect.\,\ref{subsec:notable} and mark them
	in Table\,\ref{tab:wcpar}, meaning that their derived parameters have to be taken with care. 
	For the remaining objects in our sample, we assume that they are either single or that their 
	companion contributes only negligibly to the visual spectrum, thus not significantly affecting the derived
	parameters.
	
	The structure of this paper is as follows: In Sect.\,\ref{sec:distances}, we briefly discuss the issue of converting parallaxes into distances and
	how the other updated quantities are calculated. The following Sect.\,\ref{sec:results} contains the results,
	sorted into subsections discussing the various aspects such as luminosities, mass-loss rates, or the absolute
	magnitudes of the sample stars. In Sect.\,\ref{sec:evol}, we put the revised results into an evolutionary
	context before eventually drawing the conclusions in Sect.\,\ref{sec:conclusions}. In a subsequent paper, 
	we will perform a similar study for the Galactic WN stars.
	
\section{Impact of revised distances}
  \label{sec:distances}

  In principle the distance $d$ to a particular object is the
	inverse of its parallax $\varpi$. Unfortunately, this only holds for the ``true'' parallax,
	meaning if the measurement of $\varpi$ would have no or negligible errors. An inspection of $\sigma_\varpi$, the standard deviation of $\varpi$, 
	also denoted as ``standard error'' in the Gaia data archive, reveals that while there are $12$
	sources for which the relative error $\sigma_\varpi/\varpi$ is less than $0.1$, most parallaxes have a
	larger uncertainty. Moreover, $\varpi$
	itself is not a direct measurement, but the result of interpreting the observed change of direction to a source 
	with the help of a time-dependent model describing both the motion of the actual source through space as well as the motion of
	the observational instrument, that is in this case Gaia \citep[see, e.g.,][]{Lindegren+2016}, around the sun.
	Thus the determined values of $\varpi$ are already the result from a fit procedure which,
	in the case of noisy observations, can lead to unphysical negative parallaxes. Even for positive
	values, the simple inversion of $d = \varpi^{-1}$ provides a poor distance estimate for $\sigma_\varpi/\varpi > 0.1$,
	as discussed in detail in \cite{Bailer-Jones2015} and, more recently, \citet{Luri+2018}. 
		
	For updating the distances of our Galactic WC and WO star sample, we thus decided to use the
	distances by \citet{Bailer-Jones+2018}. These are calculated from the Gaia DR2 parallaxes with the
	help of a Bayesian approach. They assume a prior with an exponentially decreasing space density in distance
	with a length-scale parameter $\ell$ that depends on Galactic latitude and longitude. Assuming a Gaussian likelihood
	for the parallax $\varpi$ as motivated by \citet{Lindegren+2018}, the distances given by \citet{Bailer-Jones+2018} 
	are then obtained by the mode of the posterior, which is derived by solving a third-order equation depending on $\ell$.
	The uncertainties for the distance are defined by the lower and upper boundaries of the highest density interval around
	the mode of the posterior with probability $p = 0.6827$. For a Gaussian posterior, these would correspond to $\pm 1\sigma$.
	To quantify the impact of these distance uncertainties, we will list corresponding error margins in Table\,\ref{tab:wcpar} on
	all quantities directly depending on the distance.
	
	In this work, we purely focus on the impact of distances and thus do not update any of the atmosphere models
	applied in \citet{Sander+2012}. This means in particular that the obtained temperature $T_\ast$ and the 
	so-called ``transformed radius''  
  \begin{equation}
  \label{eq:rt}
 R_{\mathrm{t}} = R_{\ast} \left[ \frac{\varv_{\infty}}{2500\,\mathrm{km/s}} 
 \left/ \frac{\dot{M} \sqrt{D} }{10^{-4} M_{\odot}/\mathrm{yr}} \right. 
 \right]^{\frac{2}{3}}
  \end{equation} 
  \citep{SHW1989,HK1998} remain the same. In Eq.\,(\ref{eq:rt}), $R_\ast$ denotes the stellar radius which is
	also the inner boundary of the unterlying atmosphere models defined at a Rosseland continuum optical depth 
	of $\tau = 20$. $\dot{M}$ denotes the mass-loss rate, $\varv_\infty$ the terminal wind velocity and $D$ is
	the clumping factor as introduced in \citet{HK1998}.
	Using the definition of the distance modulus
	\begin{equation}
	  \label{eq:dm}
		\text{D.M.} = 5 \log \frac{d}{10\,\text{pc}}\text{,}
  \end{equation}
	we can compare the new distances directly to those applied in \citet{Sander+2012}. With the help of the
	Stefan-Boltzmann equation
	\begin{equation}
	  \label{eq:lrt}
	  L = 4 \pi R_\ast^2 \sigma_\text{SB} T_\ast^4\text{,}
	\end{equation}
	connecting $R_\ast$ and $T_\ast$ to the stellar luminosity $L$, as well as Eq.\,(\ref{eq:rt}), we can immediately obtain
	\begin{align}
	  \Delta\log L &= 0.4 \cdot \Delta\text{D.M.} \\
    \Delta\log R_\ast &= 0.5 \cdot \Delta\log L \\
		\label{eq:mdotcalc}
    \Delta\log \dot{M} &= 0.75 \cdot \Delta\log L
	\end{align}
	with $\Delta$ denoting the difference between the new values based on the Gaia parallaxes and the old
	values from the 2012 paper.
	

\begin{figure*}
   \centering
  \includegraphics[width=17cm]{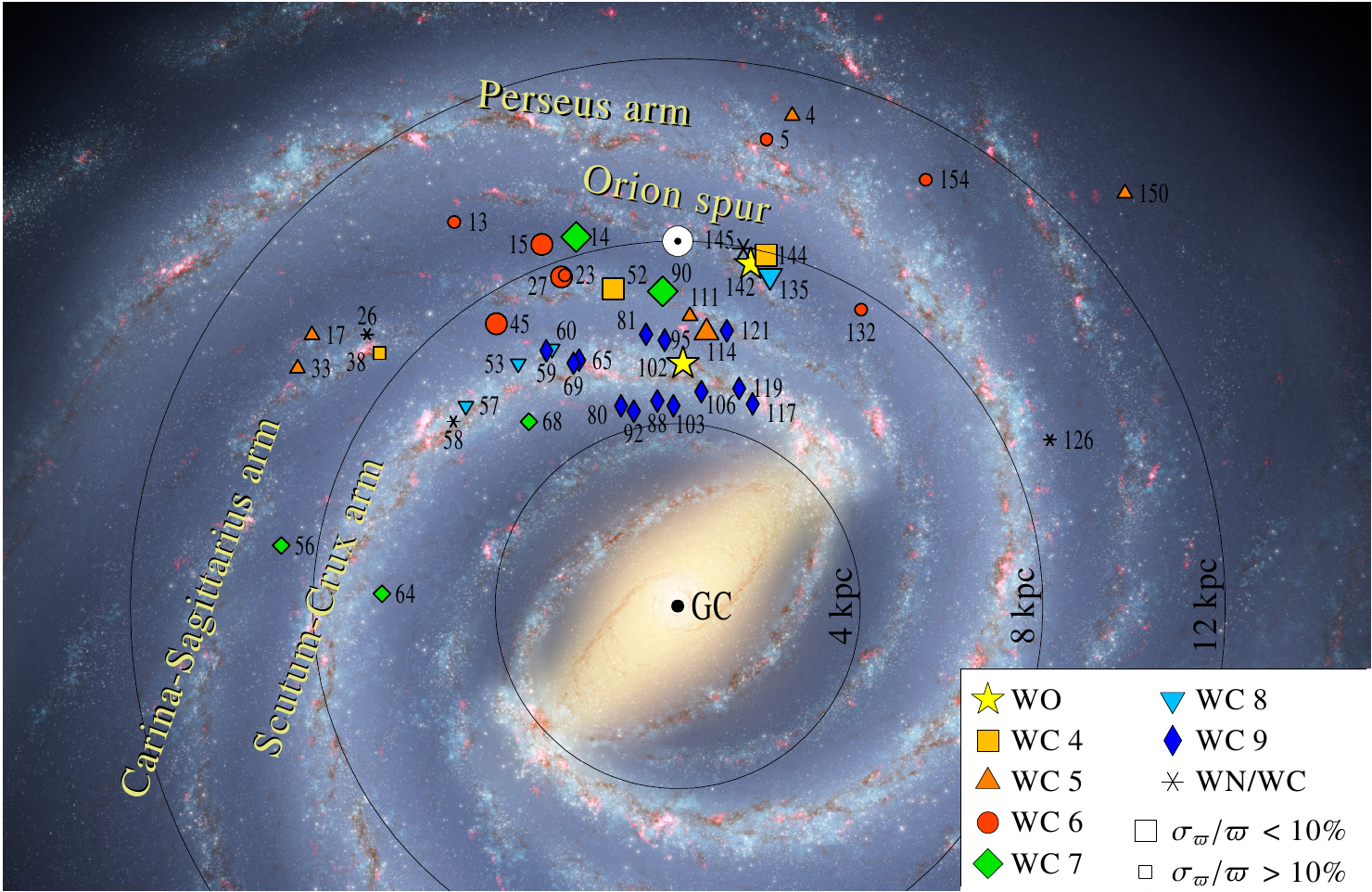}
  \caption{Positions in the Galactic plane of the Galactic WC and WO sample with the positions of the sun ($\odot$) and the Galactic Center (GC) indicated. 
	         The different symbols depict the various subtypes and larger symbols denote lower parallax errors as explained in Fig.\,\ref{fig:hrd}.
					 For illustration purposes, an artist's impression of the Milky Way has been put in the background with certain spiral arms highlighted. 
					 Background image credit: NASA/JPL-Caltech/ESO/R. Hurt}\label{fig:galwcpos}
\end{figure*}   

  Utilizing the new distances and the Galactic coordinates, we can plot the locations of the sample WC and WO stars projected on the Galactic
	plane as depicted in Fig.\,\ref{fig:galwcpos}. Though the background is an artist's impression, the spiral arm structure is roughly based on
	our current knowledge of the Milky Way structure. As expected, most WC stars and both WO stars are located in the spiral arms. For a more
	in-depth discussion about the spatial distribution of the Galactic WR stars, although not yet based on the Gaia DR2 results, we 
	refer to \citet{RossloweCrowther2015}.

\section{Results}
  \label{sec:results}

Applying the methods and equations outlined above, we obtain the new distances as well as the dependent
quantities, such as luminosities and mass-loss rates. After reviewing the errors, this section presents the results
and puts them into context, checks the validity of an $M_\varv$ calibration per subtype, and
discusses noteworthy outcomes for a few particular objects. 

\begin{table*}[p]
  \caption{Parameters of the Galactic single WC stars}
  \label{tab:wcpar}
  \centering
	\renewcommand{\arraystretch}{1.3}
  \small

  \begin{tabular}{r l r c l l l rcr r r l r r}
      \hline
      \hline
\multicolumn{1}{c}{WR} & \multicolumn{1}{l}{Subtype} & \multicolumn{1}{c}{$T_{*}$\tablefootmark{(a)}\!\!} & \multicolumn{1}{c}{$\log R_{\mathrm{t}}$\tablefootmark{(a)}\!\!\!\!} & \multicolumn{1}{c}{$\varv_{\infty}$\tablefootmark{(a)}\!\!\!\!} & \multicolumn{1}{c}{$X_\text{He}$\tablefootmark{(a)}\!\!\!\!} & \multicolumn{1}{c}{$E_{b-\varv}$\tablefootmark{(a)}\!\!\!\!} & \multicolumn{1}{c}{D.M.\tablefootmark{(b)}\!\!\!\!} & \multicolumn{1}{c}{\!\!\rule[0mm]{0mm}{3mm}\!\!} & \multicolumn{1}{c}{$M_{\varv}$} & \multicolumn{1}{c}{$R_{*}$} & \multicolumn{1}{c}{$\log\dot{M}$} & \multicolumn{1}{c}{$\log L$} & \multicolumn{1}{c}{$M$\tablefootmark{(c)}} & \multicolumn{1}{c}{$\eta$} \\
& & \multicolumn{1}{c}{[kK]} & \multicolumn{1}{c}{[$\mathrm{R}_{\odot}$]} & \multicolumn{1}{c}{[km/s]\!\!} &\!\!\!\!& \multicolumn{1}{c}{[mag]} & \multicolumn{1}{c}{[mag]} & & \multicolumn{1}{c}{[mag]} & \multicolumn{1}{c}{[$\mathrm{R}_{\odot}$]} & \multicolumn{1}{c}{[$\mathrm{M}_{\odot}/\mathrm{yr}$]} & \multicolumn{1}{c}{[$\mathrm{L}_{\odot}$]} & \multicolumn{1}{c}{[$\mathrm{M}_{\odot}$]} & \\
    \hline
102 & WO2\!\!\!\! & $200$ & $0.40$ & $5000$ & $0.30$ & $1.08$ & $12.14^{+0.17}_{-0.16}$ & $\!\!\rightarrow\!\!$ & $-1.46$ & $0.52$ & $-5.23^{+0.05}_{-0.05}$ & $5.58^{+0.07}_{-0.06}$ & $16.1^{+1.7}_{-1.4}$ & $3.8$ \\
142 & WO2\!\!\!\! & $200$ & $0.30$ & $5000$ & $0.30$ & $1.43$ & $11.09^{+0.10}_{-0.10}$ & $\!\!\rightarrow\!\!$ & $-3.13$ & $0.80$ & $-4.80^{+0.03}_{-0.03}$ & $5.96^{+0.04}_{-0.04}$ & $28.6^{+2.1}_{-1.8}$ & $4.3$ \\
38 & WC4\rule[0mm]{0mm}{5mm}\!\!\!\! & $126$ & $0.10$ & $3200$ & $0.55$ & $1.11$ & $14.22^{+0.35}_{-0.31}$ & $\!\!\rightarrow\!\!$ & $-3.36$ & $0.85$ & $-4.66^{+0.10}_{-0.09}$ & $5.21^{+0.14}_{-0.12}$ & $10.4^{+1.9}_{-1.4}$ & $21.6$ \\
52 & WC4\!\!\!\! & $112$ & $0.20$ & $3225$ & $0.35$ & $0.56$ & $11.22^{+0.15}_{-0.14}$ & $\!\!\rightarrow\!\!$ & $-3.21$ & $0.92$ & $-4.75^{+0.04}_{-0.04}$ & $5.07^{+0.06}_{-0.06}$ & $8.5^{+0.6}_{-0.5}$ & $24.1$ \\
144 & WC4\!\!\!\! & $112$ & $0.20$ & $3500$ & $0.35$ & $1.60$ & $11.24^{+0.21}_{-0.19}$ & $\!\!\rightarrow\!\!$ & $-3.28$ & $1.06$ & $-4.62^{+0.06}_{-0.06}$ & $5.20^{+0.08}_{-0.08}$ & $9.9^{+1.1}_{-0.9}$ & $26.4$ \\
4 & WC5\rule[0mm]{0mm}{5mm}\!\!\!\! & $79$ & $0.50$ & $2528$ & $0.55$ & $0.60$ & $12.85^{+0.35}_{-0.31}$ & $\!\!\rightarrow\!\!$ & $-5.14$ & $3.85$ & $-4.37^{+0.11}_{-0.09}$ & $5.71^{+0.14}_{-0.12}$ & $20.0^{+4.6}_{-3.2}$ & $10.3$ \\
17 & WC5\!\!\!\! & $79$ & $0.50$ & $2231$ & $0.55$ & $0.31$ & $14.59^{+0.65}_{-0.57}$ & $\!\!\rightarrow\!\!$ & $-5.60$ & $4.01$ & $-4.40^{+0.20}_{-0.17}$ & $5.74^{+0.26}_{-0.23}$ & $21.0^{+10.5}_{-5.7}$ & $7.8$ \\
33 & WC5\!\!\!\! & $79$ & $0.50$ & $3342$ & $0.55$ & $0.60$ & $14.72^{+0.49}_{-0.44}$ & $\!\!\rightarrow\!\!$ & $-4.77$ & $3.25$ & $-4.36^{+0.15}_{-0.13}$ & $5.56^{+0.20}_{-0.18}$ & $16.3^{+5.2}_{-3.4}$ & $19.5$ \\
111 & WC5\!\!\!\! & $89$ & $0.40$ & $2398$ & $0.55$ & $0.34$ & $11.10^{+0.31}_{-0.27}$ & $\!\!\rightarrow\!\!$ & $-4.26$ & $2.10$ & $-4.64^{+0.09}_{-0.08}$ & $5.39^{+0.12}_{-0.11}$ & $13.0^{+2.3}_{-1.7}$ & $11.0$ \\
114 & WC5\!\!\!\! & $79$ & $0.50$ & $3200$ & $0.55$ & $1.35$ & $11.61^{+0.18}_{-0.17}$ & $\!\!\rightarrow\!\!$ & $-4.19$ & $2.68$ & $-4.51^{+0.05}_{-0.05}$ & $5.39^{+0.07}_{-0.07}$ & $13.1^{+1.3}_{-1.0}$ & $19.7$ \\
150 & WC5\!\!\!\! & $89$ & $0.40$ & $3000$ & $0.55$ & $0.80$ & $14.97^{+0.41}_{-0.38}$ & $\!\!\rightarrow\!\!$ & $-5.26$ & $3.59$ & $-4.19^{+0.12}_{-0.11}$ & $5.86^{+0.16}_{-0.15}$ & $24.8^{+7.4}_{-5.0}$ & $13.3$ \\
5 & WC6\rule[0mm]{0mm}{5mm}\!\!\!\! & $79$ & $0.50$ & $2120$ & $0.55$ & $0.85$ & $12.36^{+0.23}_{-0.21}$ & $\!\!\rightarrow\!\!$ & $-4.61$ & $3.12$ & $-4.59^{+0.07}_{-0.06}$ & $5.53^{+0.09}_{-0.08}$ & $15.5^{+2.1}_{-1.6}$ & $7.9$ \\
13 & WC6\!\!\!\! & $79$ & $0.50$ & $2000$ & $0.55$ & $1.21$ & $13.46^{+0.36}_{-0.32}$ & $\!\!\rightarrow\!\!$ & $-4.64$ & $3.28$ & $-4.58^{+0.11}_{-0.10}$ & $5.57^{+0.14}_{-0.13}$ & $16.5^{+3.7}_{-2.6}$ & $6.9$ \\
15 & WC6\!\!\!\! & $79$ & $0.50$ & $2675$ & $0.55$ & $1.23$ & $12.37^{+0.19}_{-0.17}$ & $\!\!\rightarrow\!\!$ & $-5.69$ & $5.31$ & $-4.14^{+0.06}_{-0.05}$ & $5.99^{+0.08}_{-0.07}$ & $30.6^{+4.1}_{-3.2}$ & $9.7$ \\
23 & WC6\!\!\!\! & $79$ & $0.50$ & $2342$ & $0.55$ & $0.55$ & $12.07^{+0.22}_{-0.20}$ & $\!\!\rightarrow\!\!$ & $-4.91$ & $3.43$ & $-4.49^{+0.07}_{-0.06}$ & $5.61^{+0.09}_{-0.08}$ & $17.4^{+2.3}_{-1.8}$ & $9.2$ \\
27 & WC6\!\!\!\! & $79$ & $0.50$ & $2100$ & $0.55$ & $1.40$ & $12.13^{+0.19}_{-0.18}$ & $\!\!\rightarrow\!\!$ & $-3.92$ & $2.35$ & $-4.78^{+0.06}_{-0.05}$ & $5.28^{+0.08}_{-0.07}$ & $11.3^{+1.1}_{-0.9}$ & $9.0$ \\
45 & WC6\!\!\!\! & $79$ & $0.50$ & $2200$ & $0.55$ & $1.44$ & $13.20^{+0.16}_{-0.15}$ & $\!\!\rightarrow\!\!$ & $-4.30$ & $2.87$ & $-4.63^{+0.05}_{-0.04}$ & $5.45^{+0.06}_{-0.06}$ & $14.1^{+1.2}_{-1.0}$ & $9.0$ \\
132 & WC6\!\!\!\! & $71$ & $0.60$ & $2400$ & $0.55$ & $1.15$ & $13.16^{+0.27}_{-0.24}$ & $\!\!\rightarrow\!\!$ & $-4.38$ & $3.15$ & $-4.67^{+0.08}_{-0.07}$ & $5.35^{+0.11}_{-0.10}$ & $12.4^{+1.8}_{-1.4}$ & $11.3$ \\
154 & WC6\!\!\!\! & $79$ & $0.50$ & $2300$ & $0.55$ & $0.78$ & $13.74^{+0.41}_{-0.36}$ & $\!\!\rightarrow\!\!$ & $-5.59$ & $4.88$ & $-4.26^{+0.12}_{-0.11}$ & $5.91^{+0.16}_{-0.14}$ & $27.2^{+8.3}_{-5.3}$ & $7.6$ \\
14 & WC7\rule[0mm]{0mm}{5mm}\!\!\!\! & $71$ & $0.60$ & $2194$ & $0.55$ & $0.65$ & $11.74^{+0.16}_{-0.15}$ & $\!\!\rightarrow\!\!$ & $-5.37$ & $5.16$ & $-4.39^{+0.05}_{-0.05}$ & $5.78^{+0.07}_{-0.06}$ & $22.1^{+2.3}_{-1.9}$ & $7.3$ \\
56 & WC7\!\!\!\! & $71$ & $0.60$ & $2009$ & $0.55$ & $0.70$ & $15.20^{+0.41}_{-0.37}$ & $\!\!\rightarrow\!\!$ & $-4.13$ & $3.09$ & $-4.76^{+0.12}_{-0.11}$ & $5.33^{+0.16}_{-0.15}$ & $12.1^{+2.8}_{-2.0}$ & $7.9$ \\
64 & WC7\!\!\!\! & $71$ & $0.60$ & $1700$ & $0.55$ & $1.20$ & $15.02^{+0.47}_{-0.42}$ & $\!\!\rightarrow\!\!$ & $-4.35$ & $2.87$ & $-4.88^{+0.14}_{-0.13}$ & $5.27^{+0.19}_{-0.17}$ & $11.2^{+3.0}_{-2.1}$ & $5.9$ \\
68 & WC7\!\!\!\! & $71$ & $0.60$ & $2100$ & $0.55$ & $1.40$ & $13.55^{+0.38}_{-0.33}$ & $\!\!\rightarrow\!\!$ & $-5.16$ & $4.96$ & $-4.44^{+0.11}_{-0.10}$ & $5.74^{+0.15}_{-0.13}$ & $21.0^{+5.4}_{-3.6}$ & $6.8$ \\
90 & WC7\!\!\!\! & $71$ & $0.60$ & $2053$ & $0.55$ & $0.40$ & $10.31^{+0.15}_{-0.14}$ & $\!\!\rightarrow\!\!$ & $-4.99$ & $4.00$ & $-4.59^{+0.04}_{-0.04}$ & $5.55^{+0.06}_{-0.06}$ & $16.2^{+1.4}_{-1.2}$ & $7.3$ \\
53 & WC8d\rule[0mm]{0mm}{5mm}\!\!\!\! & $50$ & $0.90$ & $1800$ & $0.55$ & $0.75$ & $13.22^{+0.33}_{-0.29}$ & $\!\!\rightarrow\!\!$ & $-5.41$ & $7.77$ & $-4.66^{+0.10}_{-0.09}$ & $5.52^{+0.13}_{-0.12}$ & $15.5^{+3.1}_{-2.2}$ & $5.8$ \\
57 & WC8\!\!\!\! & $63$ & $0.70$ & $1787$ & $0.55$ & $0.38$ & $13.85^{+0.54}_{-0.46}$ & $\!\!\rightarrow\!\!$ & $-5.61$ & $6.36$ & $-4.50^{+0.16}_{-0.14}$ & $5.75^{+0.22}_{-0.18}$ & $21.2^{+8.4}_{-4.8}$ & $4.9$ \\
60 & WC8\!\!\!\! & $63$ & $0.70$ & $2300$ & $0.55$ & $1.45$ & $12.80^{+0.31}_{-0.27}$ & $\!\!\rightarrow\!\!$ & $-5.49$ & $6.77$ & $-4.35^{+0.09}_{-0.08}$ & $5.80^{+0.12}_{-0.11}$ & $23.0^{+4.8}_{-3.4}$ & $8.0$ \\
135 & WC8\!\!\!\! & $63$ & $0.60$ & $1343$ & $0.75$ & $0.40$ & $11.50^{+0.15}_{-0.14}$ & $\!\!\rightarrow\!\!$ & $-4.78$ & $4.24$ & $-4.73^{+0.04}_{-0.04}$ & $5.40^{+0.06}_{-0.06}$ & $13.6^{+1.1}_{-0.9}$ & $4.9$ \\
59\tablefootmark{(d)}\!\!\!\!\!\! & WC9d\rule[0mm]{0mm}{5mm}\!\!\!\! & $40$ & $1.00$ & $1300$ & $0.55$ & $2.00$ & $12.87^{+0.33}_{-0.29}$ & $\!\!\rightarrow\!\!$ & $-7.31$ & $15.89$ & $-4.48^{+0.10}_{-0.09}$ & $5.76^{+0.13}_{-0.12}$ & $21.4^{+4.7}_{-3.3}$ & $3.7$ \\
65\tablefootmark{(d)}\!\!\!\!\!\! & WC9d\!\!\!\! & $40$ & $1.00$ & $1300$ & $0.55$ & $2.00$ & $12.65^{+0.43}_{-0.37}$ & $\!\!\rightarrow\!\!$ & $-7.09$ & $18.50$ & $-4.38^{+0.13}_{-0.11}$ & $5.89^{+0.17}_{-0.15}$ & $26.2^{+8.5}_{-5.2}$ & $3.4$ \\
69 & WC9d\!\!\!\! & $40$ & $1.00$ & $1089$ & $0.55$ & $0.55$ & $12.73^{+0.29}_{-0.26}$ & $\!\!\rightarrow\!\!$ & $-5.51$ & $9.77$ & $-4.87^{+0.09}_{-0.08}$ & $5.33^{+0.12}_{-0.10}$ & $12.1^{+1.9}_{-1.4}$ & $3.3$ \\
80 & WC9d\!\!\!\! & $45$ & $0.90$ & $1600$ & $0.55$ & $1.80$ & $12.91^{+0.65}_{-0.53}$ & $\!\!\rightarrow\!\!$ & $-5.88$ & $6.89$ & $-4.79^{+0.20}_{-0.16}$ & $5.24^{+0.26}_{-0.21}$ & $10.8^{+4.2}_{-2.4}$ & $7.5$ \\
81 & WC9\!\!\!\! & $45$ & $0.80$ & $1600$ & $0.55$ & $1.50$ & $11.67^{+0.31}_{-0.27}$ & $\!\!\rightarrow\!\!$ & $-5.44$ & $7.08$ & $-4.62^{+0.09}_{-0.08}$ & $5.26^{+0.12}_{-0.11}$ & $11.1^{+1.8}_{-1.4}$ & $10.4$ \\
88 & WC9/WN8\!\!\!\! & $40$ & $1.00$ & $1500$ & $0.55$ & $1.40$ & $12.74^{+0.41}_{-0.35}$ & $\!\!\rightarrow\!\!$ & $-5.78$ & $11.73$ & $-4.62^{+0.12}_{-0.11}$ & $5.49^{+0.16}_{-0.14}$ & $14.9^{+3.7}_{-2.5}$ & $5.7$ \\
92 & WC9\!\!\!\! & $45$ & $0.80$ & $1121$ & $0.55$ & $0.52$ & $12.93^{+0.52}_{-0.43}$ & $\!\!\rightarrow\!\!$ & $-4.50$ & $4.97$ & $-5.00^{+0.16}_{-0.13}$ & $4.95^{+0.21}_{-0.17}$ & $7.7^{+2.1}_{-1.3}$ & $6.1$ \\
95 & WC9d\!\!\!\! & $45$ & $0.90$ & $1900$ & $0.55$ & $1.74$ & $11.69^{+0.37}_{-0.32}$ & $\!\!\rightarrow\!\!$ & $-5.25$ & $6.86$ & $-4.71^{+0.11}_{-0.09}$ & $5.23^{+0.15}_{-0.13}$ & $10.7^{+2.1}_{-1.5}$ & $10.8$ \\
103 & WC9d\!\!\!\! & $45$ & $0.80$ & $1190$ & $0.55$ & $0.52$ & $12.79^{+0.61}_{-0.49}$ & $\!\!\rightarrow\!\!$ & $-6.06$ & $9.31$ & $-4.56^{+0.18}_{-0.15}$ & $5.50^{+0.24}_{-0.19}$ & $15.0^{+6.0}_{-3.3}$ & $5.1$ \\
106 & WC9d\!\!\!\! & $45$ & $0.80$ & $1100$ & $0.55$ & $1.20$ & $12.62^{+0.38}_{-0.33}$ & $\!\!\rightarrow\!\!$ & $-5.36$ & $6.81$ & $-4.80^{+0.11}_{-0.10}$ & $5.23^{+0.15}_{-0.13}$ & $10.6^{+2.2}_{-1.5}$ & $5.1$ \\
117 & WC9d\!\!\!\! & $56$ & $0.60$ & $2000$ & $0.55$ & $1.56$ & $12.97^{+0.70}_{-0.57}$ & $\!\!\rightarrow\!\!$ & $-5.19$ & $5.12$ & $-4.44^{+0.21}_{-0.17}$ & $5.36^{+0.28}_{-0.23}$ & $12.5^{+5.5}_{-3.0}$ & $15.5$ \\
119 & WC9d\!\!\!\! & $45$ & $0.80$ & $1300$ & $0.55$ & $0.90$ & $12.72^{+0.62}_{-0.51}$ & $\!\!\rightarrow\!\!$ & $-3.91$ & $3.70$ & $-5.13^{+0.19}_{-0.15}$ & $4.70^{+0.25}_{-0.20}$ & $5.8^{+1.8}_{-1.1}$ & $9.6$ \\
121 & WC9d\!\!\!\! & $45$ & $0.80$ & $1100$ & $0.55$ & $1.40$ & $11.75^{+0.24}_{-0.21}$ & $\!\!\rightarrow\!\!$ & $-5.08$ & $6.35$ & $-4.85^{+0.07}_{-0.06}$ & $5.16^{+0.09}_{-0.09}$ & $9.9^{+1.2}_{-0.9}$ & $5.3$ \\
26 & WN7/WCE\rule[0mm]{0mm}{5mm}\!\!\!\! & $79$ & $0.60$ & $2700$ & $0.78$ & $1.25$ & $14.26^{+0.27}_{-0.24}$ & $\!\!\rightarrow\!\!$ & $-4.74$ & $3.94$ & $-4.29^{+0.08}_{-0.07}$ & $5.73^{+0.11}_{-0.10}$ & $21.1^{+3.5}_{-2.6}$ & $12.7$ \\
58 & WN4/WCE\!\!\!\! & $79$ & $0.50$ & $1600$ & $0.98$ & $0.55$ & $14.00^{+0.46}_{-0.40}$ & $\!\!\rightarrow\!\!$ & $-3.34$ & $1.61$ & $-4.95^{+0.14}_{-0.12}$ & $4.95^{+0.19}_{-0.16}$ & $8.4^{+1.9}_{-1.3}$ & $9.9$ \\
126 & WC5/WN\!\!\!\! & $63$ & $1.20$ & $2000$ & $0.75$ & $0.95$ & $14.83^{+0.49}_{-0.43}$ & $\!\!\rightarrow\!\!$ & $-5.43$ & $9.23$ & $-4.96^{+0.15}_{-0.13}$ & $6.07^{+0.20}_{-0.17}$ & $35.9^{+15.2}_{-8.6}$ & $0.9$ \\
145\tablefootmark{(e)}\!\!\!\!\!\! & WN7/WCE\!\!\!\! & $50$ & $0.90$ & $1440$ & $0.98$ & $1.86$ & $10.82^{+0.14}_{-0.13}$ & $\!\!\rightarrow\!\!$ & $-5.90$ & $8.59$ & $-4.49^{+0.04}_{-0.04}$ & $5.61^{+0.06}_{-0.05}$ & $18.3^{+1.4}_{-1.2}$ & $5.6$ \\
      \hline
  \end{tabular}
	\vspace{-5pt}
  \tablefoot{
      \tablefoottext{a}{Unchanged from \citet{Sander+2012}}
      \tablefoottext{b}{All distance moduli based on Gaia DR2 parallaxes converted into distances by \citet{Bailer-Jones+2018}.}
      \tablefoottext{c}{Masses are calculated from luminosity after \citet{Langer89mass} using their $M$-$L$ relation for WC stars.}
			\tablefoottext{d}{Binary candidate with possibly significant contribution from the secondary. See comments in Sect.\,\ref{subsec:notable}}
			\tablefoottext{e}{Binary system analyzed as a single source. See comments in Sect.\,\ref{subsec:notable}}
  }
\end{table*}

\subsection{Error estimates}
  \label{subsec:errors}
  
In our previous analysis of the Galactic WC stars \citep{Sander+2012}, the spectroscopic parameters ($T_\ast, R_\text{t}$)
were determined by assigning the best model from a grid of WC atmospheres reproducing the normalized line spectrum of the available observations.
The uncertainties of the assigned spectroscopic parameters roughly correspond to the span between neighboring models in the grid. For dense winds,
there is a certain parameter degeneracy \citep{HGK2003}.
The luminosities were derived from reproducing the spectral energy distribution (SED), which requires an input of
the reddening parameter $E_{b-\varv}$ and the distance. While the former could be determined quite accurately with
an uncertainty of about $\pm0.02\,$mag, the latter depended on the often uncertain membership of the
particular object to a cluster or association and the uncertainty of their distance themselves. Thus, the distance moduli
had an estimated error margin of about $\pm0.75\,$mag, outweighing all other uncertainties that enter the luminosity
estimate. 
In this work, we focus only on the distances and their uncertainty.
As mentioned in Sect.\,\ref{sec:distances}, the distances calculated from the Gaia DR2 parallaxes by \citet{Bailer-Jones+2018}
come with uncertainty estimates intrinsic to their Bayesian method, defined by the the lower and upper boundaries of an 
asymmetric confidence interval that corresponds to one standard deviation in the case of the distance distribution being described by a Gaussian.
With these uncertainties given, we calculate the resulting uncertainties for the distance modulus, the luminosity and the mass-loss rate
according to the Eqs.\,(\ref{eq:dm}) to (\ref{eq:mdotcalc}). We are aware that these would also propagate into the new values of
$R_\ast$, but this value listed in Table\,\ref{tab:wcpar} serves more as a reference quantity between different types of WC stars. 

We calculate the stellar masses by inverting Eq.\,(19) from \citet{Langer89mass} that connects masses and luminosities of WC stars. Unlike hydrogen-free
WN stars, which should be from a theoretical standpoint relatively close to ideal He main sequence stars, WC stars already have a significant
carbon and oxygen fraction at the surface. Consequently, helium is depleted and -- unless the star is fully mixed -- even more so in the inner layers
down to the core. Therefore, the helium (surface) mass fraction enters this relation as an additional parameter, although at a moderate degree.
As an example, for a WC star with $\log L/L_\odot = 5.8$, the masses range from $21.7\,M_\odot$ in the case of no helium at all to $24.1\,M_\odot$ in the
case of pure helium, which would of course technically no longer be classified as a WC star. Assuming that the helium mass fraction lies in a 
realistic range of $X_\text{He} = 0.3\dots0.8$, the previous example would lead to an uncertainty of 
$\sim\!1\,M_\odot$. This is much lower compared to the uncertainty in the luminosity and thus is ignored in our error
estimate given in Table\,\ref{tab:wcpar}.

\begin{figure}[tbh!]
  \resizebox{\hsize}{!}{\includegraphics{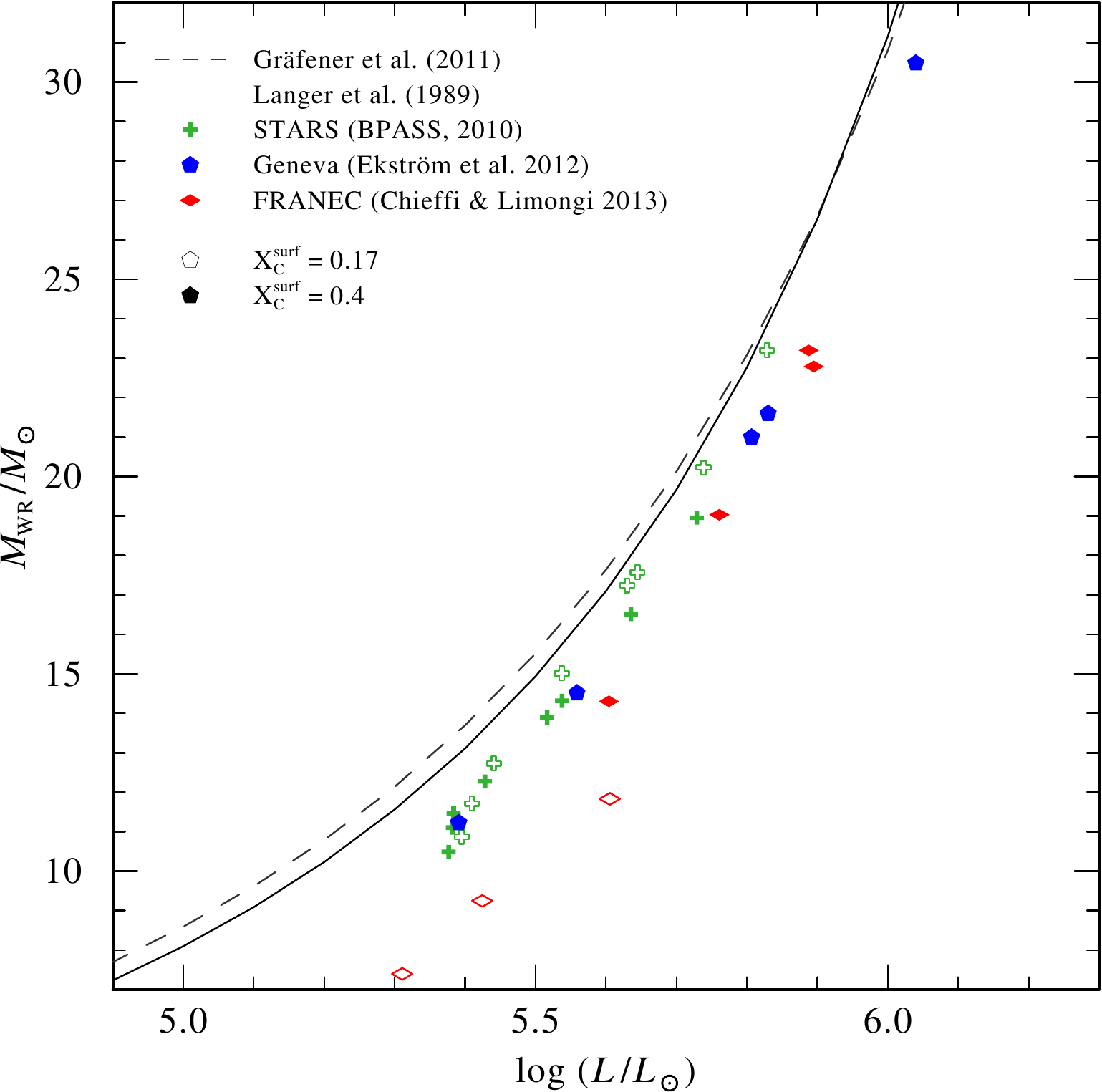}}
  \caption{Comparison of the mass-luminosity relations from \citet{Langer89mass} for WC stars and \citet{Graefener+2011} for hydrogen-free WN stars with
	         snapshots of different stellar evolution models showing Carbon surface abundances which are commonly attributed to a WC spectral appearance.}
  \label{fig:masslum}
\end{figure}

To check whether different WC masses would be derived from newer stellar evolution models with e.g. updated compositions, opacities, or nuclear energy generation rates, 
 we compare the relation from \citet{Langer89mass} with
 the location of selected stages (``snapshots'') from different stellar evolution codes in the mass-luminosity plane. This is depicted in Fig.\,\ref{fig:masslum},
 where we plot the snapshot of evolution models with carbon surface mass fractions $X_\text{C} = 0.4$ for the Geneva \citep{Ekstroem+2012} and FRANEC \citep{CL2013} codes
 which account for rotation, as well as single star models calculated with the STARS code in 2010 from BPASS \citep{ES2009}. For comparison,
 also the relation for chemically-homogeneous hydrogen-free WN stars from \citet{Graefener+2011} is shown. Since some tracks do not reach $X_\text{C} = 0.4$,
 which is the standard carbon abundance in our sample, we also show selected snapshots of $X_\text{C} = 0.17$ where a star should already have a WC-type
 spectrum (assuming of course a sufficiently high mass-loss rate). While the relation from \citet{Langer89mass} used in this work gives masses that are
 within $1\,M_\odot$ of the He-star relation, the evolution models yield a bit lower masses, typically by $1$ to $2\,M_\odot$ for the models resulting from
 the STARS and Geneva code. The FRANEC results differ a bit more from the other models, especially for the lower masses where the models do only reach $X_\text{C} = 0.17$.
 Since these values are just based on snapshots and none of the papers provide a mass-luminosity relation for WC stars, we stick with the relation from \citet{Langer89mass},
 but point out that the WC masses derived here could be slightly overestimated.

\subsection{Luminosities, temperatures and HRD positions}
  \label{subsec:lumo}
  
\begin{figure*}[tbh!]
  \sidecaption
  \includegraphics[width=12cm]{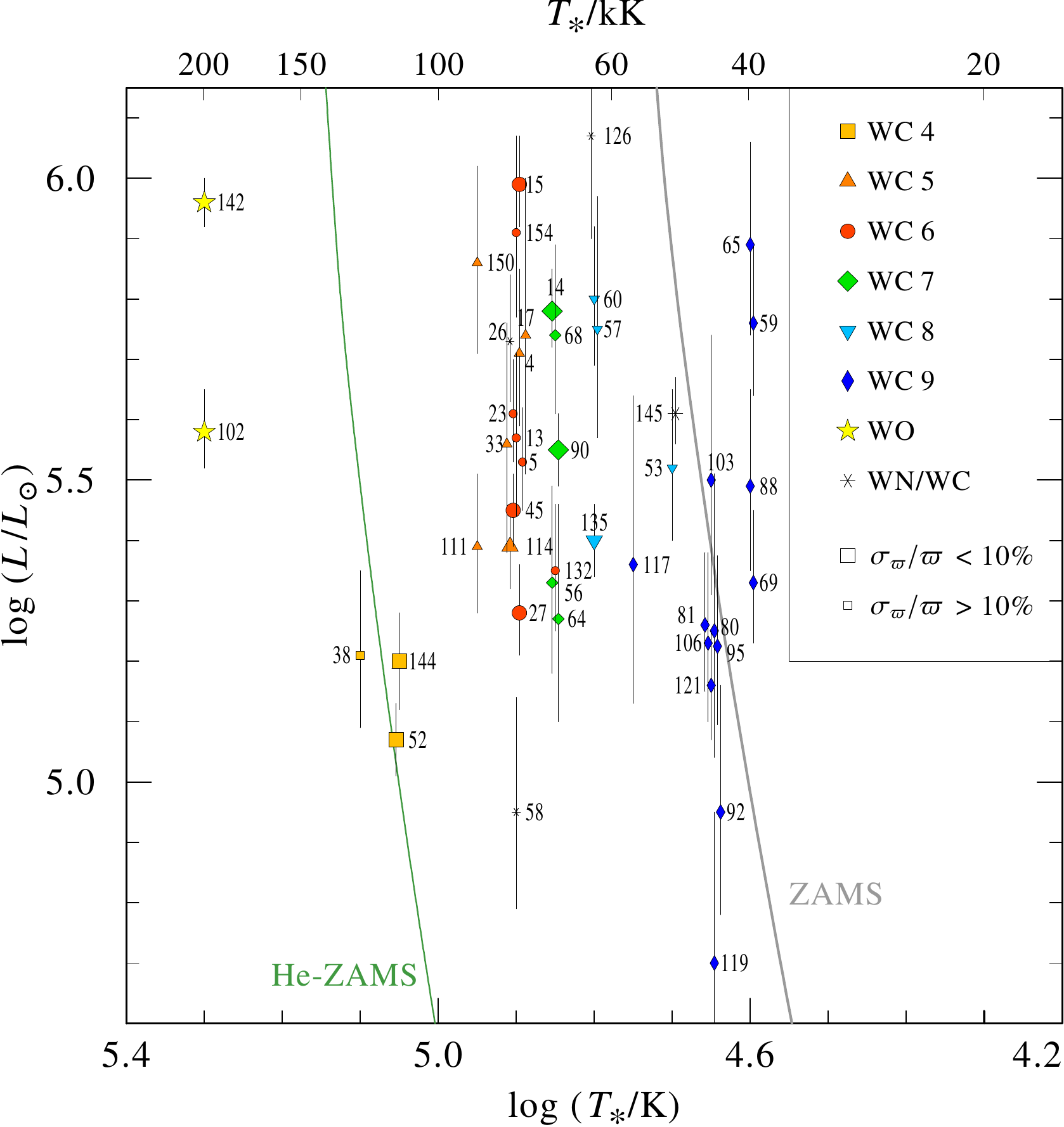}
  \caption{Hertzsprung-Russell diagram of the
Galactic WC \& WO stars with updated luminosities due to Gaia distances: The sample is taken from \citet{Sander+2012}, reduced to those with Gaia parallaxes that have a relative
error of less then $40\%$. The vertical bars denote the error in luminosity resulting from the distance uncertainty. Tiny horizontal shifts have been applied to different different
symbols for clarity. Stars with parallax errors of less than ten percent are represented by larger
symbols. The zero age main sequence \citep[ZAMS, from][here for stars without initial rotation]{Ekstroem+2012} and for pure helium stars \citep[He-ZAMS, from][]{Langer89mass} are
depicted as gray and green solid lines.}\label{fig:hrd}
\end{figure*}    

\begin{figure}[tbh!]
  \resizebox{\hsize}{!}{\includegraphics{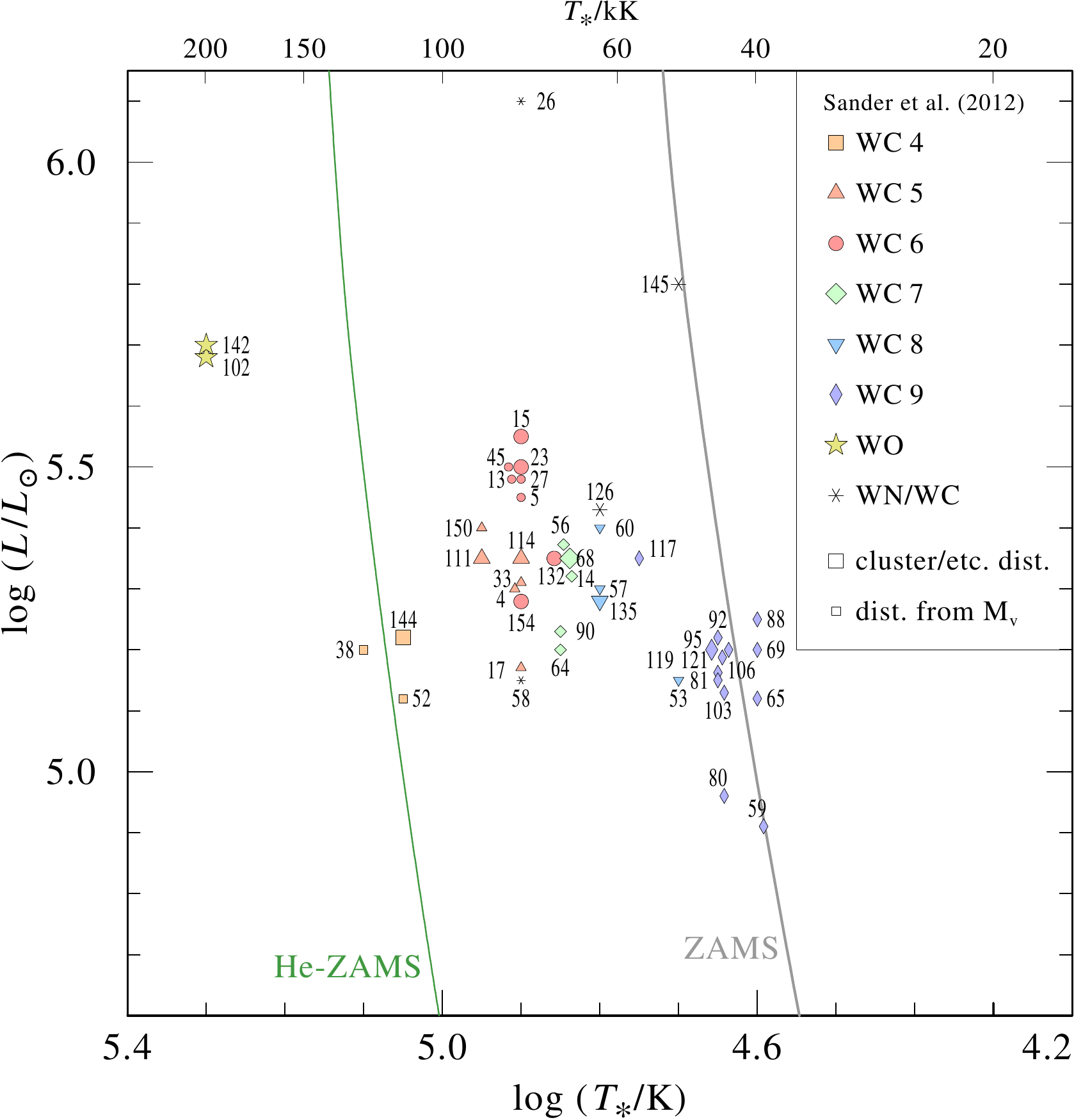}}
  \caption{Hertzsprung-Russell diagram of the Galactic WC \& WO stars with their luminosities as in the \citet{Sander+2012} paper. The larger symbols denote stars where a direct
	information for the distance, e.g. from cluster membership, was known. The smaller symbols denote stars with luminosities that were obtained via a subtype-magnitude calibration.}
  \label{fig:hrd2012}
\end{figure}

  Applying the new luminosities based on Gaia DR2 parallaxes and taking $T_\ast$ from \citet{Sander+2012}, we obtain the revised positions of the 
	Galactic WC and WO stars in the Hertzsprung-Russell diagram (HRD) as shown in Fig.\,\ref{fig:hrd}. For comparison, the HRD with the luminosities used in the 2012 paper is shown
	in Fig.\,\ref{fig:hrd2012}. While a number of stars more or less remain at their previous position, most of them change, some even quite drastically. Already at first sight it becomes
	clear that the luminosity range between $\log L/L_\odot = 5.6$ and $6.0$, which was only populated by the two WO2 stars previously, is now also covered by various stars from the subtypes
	WC5 to WC9. This previous upper luminosity limit was likely too low due to the $M_\varv$ calibration as we discuss in Sect.\,\ref{subsec:mvrel}. 
	
	With the exception of the WC4 stars, all other WC subtypes cover a broad range of luminosities. This is in particular true for the WC9 subtype, ranging over more than one order of
	magnitude in $L$ instead of just clustering around $\log L/L_\odot \approx 5.2$ in the 2012 paper. The reason for the earlier results was the fact that only one star, WR~95, had 
	a constrained distance and thus all other WC9 stars were calibrated to its magnitude. While the results based on Gaia DR2 confirm the HRD position of WR~95, they also reveal
	that there is a significant spread in luminosities for WC9 stars, independent of whether or not they are WC9d, that is showing infrared dust excess. As we discuss in Sect.\,\ref{subsec:notable}, the two most luminous WC9 stars are most likely binaries, which means that their luminosity in Table\,\ref{tab:wcpar} would be overestimated. However, even with a
	significant luminosity reduction of these two stars, for example by $0.3\,$dex, the spread in WC9 luminosities would still be about one order of magnitude.
	
	Apart from the change in the overall picture due to the revised distance, there are several noteworthy objects that significantly altered their obtained luminosities. These is discussed in Sect.\,\ref{subsec:notable}.
	
	The temperatures of the WC stars have been a constant topic of discussion in the past years. Since one can safely assume that WC stars are core-helium burning objects
	and do not have a hydrogen layer, one would expect them to appear relatively close to the zero age main sequence (ZAMS) for pure helium stars (He-ZAMS) or even hotter 
	than that as they come relatively close to the 
	theoretical concept of an evolved helium star. However, Fig.\,\ref{fig:hrd2012} shows that this is only the case for the WC4 and WO stars, while all other subtypes
	show much cooler temperatures. To properly put this into perspective, one has to be aware of the different temperatures that are referred to in the literature. 
	The primary input parameter of a PoWR model is the temperature $T_\ast$, which is defined as the effective temperature according to 
	the Stefan-Boltzmann Eq.\,(\ref{eq:lrt}) at a Rosseland continuum optical depth of $\tau = 20$. For OB stars and stars of later spectral types, an effective temperature 
	is often defined at a much lower optical depth, namely at $\tau = 2/3$, which we denote as $T_{2/3}$ here to
	avoid any confusion. For stars with dense winds such as the Galactic WR stars, the difference between $T_{2/3}$ and $T_\ast$ can be up to a factor of two or even
	more in extreme cases with $T_{2/3}$ consequently being much smaller than $T_\ast$. In particular, models with different $T_\ast$ can have the 
	same $T_{2/3}$ \citep[see, e.g., Fig.\,4 in][]{Todt+2015}, illustrating the parameter degeneracy mentioned in Sect.\,\ref{subsec:errors}. In the dense wind
	regime, models with the same $T_{2/3}$ tend to have relatively similar spectra, meaning that to a certain degree one could shift them to a different, i.e.~also higher, $T_\ast$
	\citep[cf. Sect.\,4 in][]{HG2004}. Nevertheless, tests have shown that it is not possible to obtain a satisfactory fit for all stars by shifting them to the high 
	temperatures of the He-ZAMS as the models yield too strong emission lines, but more studies on this topic are currently underway. The unanswered questions regarding optically 
	thick winds of WR stars and their consequences for the mass-loss are a major topic of current research, as demonstrated by several recent 
	publications \citep[e.g.][]{Graefener+2017,NS2018,Grassitelli+2018}.
	
	Aside from the modeling aspect, one also has to be aware of the temperature definition used to construct the He-ZAMS. In our figures, we plot the He-ZAMS from \citet{Langer89mass}, but the choice
	of the set of evolutionary models is of no particular importance since all of them provide a temperature corresponding to the hydrostatic layers, not corrected for any wind.
	Some evolutionary models try to account for the wind layers and thus also provide a ``corrected'' temperature that could be compared to $T_{2/3}$, but this correction
	is usually done in a rather approximative way. So far, only a pilot study by \citet{Groh+2014} exists where detailed atmosphere models were calculated for various points
	of the evolutionary tracks, also demonstrating that more approximate corrections seem to be way off for Wolf-Rayet stars.
	Comparing the ``uncorrected'' temperatures with $T_\ast$ is therefore still usually the best method. This works well if the atmosphere models have a (quasi-)hydrostatic 
	stratification at the definition radius of $T_\ast$, i.e. $\tau = 20$. For OB
	star models this is always the case, but for Wolf-Rayet stars this is not guaranteed by default. Especially models with dense winds can have $\varv(R_\ast) \gg 1\,$km/s, meaning
	that they would not at all be hydrostatic even in these deep layers. Fortunately, the models applied in \citet{Sander+2012} mostly have $\varv(R_\ast) = 0.1\dots1\,$km/s, and
	thus this does not seem to be an issue here, at least as long as we assume that a revision of the stellar atmosphere models would not lead to a major change. This assumption
	seems justified from our experiences with particular sources and the mostly decent changes with regards to the spectral appearance when we updated our model grids
	with the latest code version.
	
	If the temperature discrepancy cannot be explained by parameter degeneracy or different physical conditions that would prevent a proper comparison, it must reflect a more
	fundamental issue. Since for a given luminosity, the temperature $T_\ast$ directly corresponds to a radius $R_\ast$ via the Stefan-Boltzmann relation, this has also been
	termed the ``WR radius problem''. Pointing out this problem and building on ideas by \citet{Ishii+1999} and \citet{Petrovic+2006}, \citet{Graefener+2012} demonstrated 
	that by accounting for reasonable clumping factors in the observed range, there could be an inflated envelope that would essentially lead to a bending of the He-ZAMS towards
	lower temperatures. This inflation is caused by the proximity to the Eddington limit ($\Gamma_\text{e} > 0.3$) and only occurs when the outer boundary temperature 
	for the stellar structure model, which should
	roughly correspond to the temperature at the sonic point, is below $\approx 70\,$kK. Considering this limit could explain why the subtypes later than WC6 are inflated while
	the WC4 class is not. The WC5 and WC6 stars have stellar temperatures between $70$ and $100\,$kK, which would be a challenge to this picture as they seem to be inflated allthough
	they should not be according to \citet{Graefener+2012}, but 
	this discrepancy might be arbitrary as the sonic point of the models used in \citet{Sander+2012} is not really well defined due to the use of a fixed velocity stratification 
	with $\beta = 1$, which is most likely not valid in the deeper layers. Moreover, the WC5 star WR111 was modeled by \citet{GH2005} with a hydrodynamically consistent model 
	using $T_\ast = 140\,$kK, thus pointing at the possibility that with a more adequate stratification also WC5 and maybe even WC6 stars might be explained by models that 
	do not require an inflation at all.
	With the atmosphere models and the luminosities given, we can obtain the values for $\Gamma_\text{e}$ using the mass-luminosity relation for WC stars from \citet{Langer89mass}. 
	For our sample stars, the derived values for $\Gamma_\text{e}$ are between $0.2$ and $0.4$ for most of the stars with an average of $\Gamma_\text{e} = 0.3$
	for all subtypes from WC5 to WC8, while the WC9 stars have a mean of $0.23$ with a large scatter between $0.11$ (WR 119) and $0.38$ (WR 65). The likely non-inflated WC and WO stars
	have with $\Gamma_\text{e}^\text{WC4} \approx 0.2$ and $\Gamma_\text{e}^\text{WO} \approx 0.4$ very different mean values. Taking these values, not all sample stars would fulfill
	the necessary inflation condition from \citet{Graefener+2012}, but we view this only as a minor concern as the we did not perform an in-depth atmosphere analysis for each of the stars
	in \citet{Sander+2012} and relied on grid models for most objects. With the resulting uncertainty in $\Gamma_\text{e}$, most of the sample stars with subtypes later than WC4 
	are thus likely to fulfill the condition of $\Gamma_\text{e} > 0.3$.
	
	The inflation of Wolf-Rayet radii was also deduced by \citet{Glatzel2008} on the basis of numerical simulations of strange-mode instabilities evolving into the nonlinear regime
	for a Wolf-Rayet star model, which was motivated by the hydrogen-free WN star WR 123. \citet{Glatzel2008} concluded that the star inflates due to consecutive shock waves generated by 
	strange-mode instabilities, which would push the matter to larger radii and potentially even launch the stellar wind. While the latter statement would be a huge paradigm shift
	and might not be the full truth given that \citet{GH2005,GH2008} demonstrated that radiative driving plays an important role, there is a certain level of consistency between 
	\citet{Glatzel2008} and \citet{Graefener+2012} as well as \citet{GraefenerVink2013} about where inflation starts and the restriction to certain opacity regimes. We  
	come back to this issue when discussing stellar evolution in Sect.\,\ref{sec:evol}.

\subsection{Mass-loss rates}
  \label{subsec:mdot}	 

\begin{figure}
  \resizebox{\hsize}{!}{\includegraphics{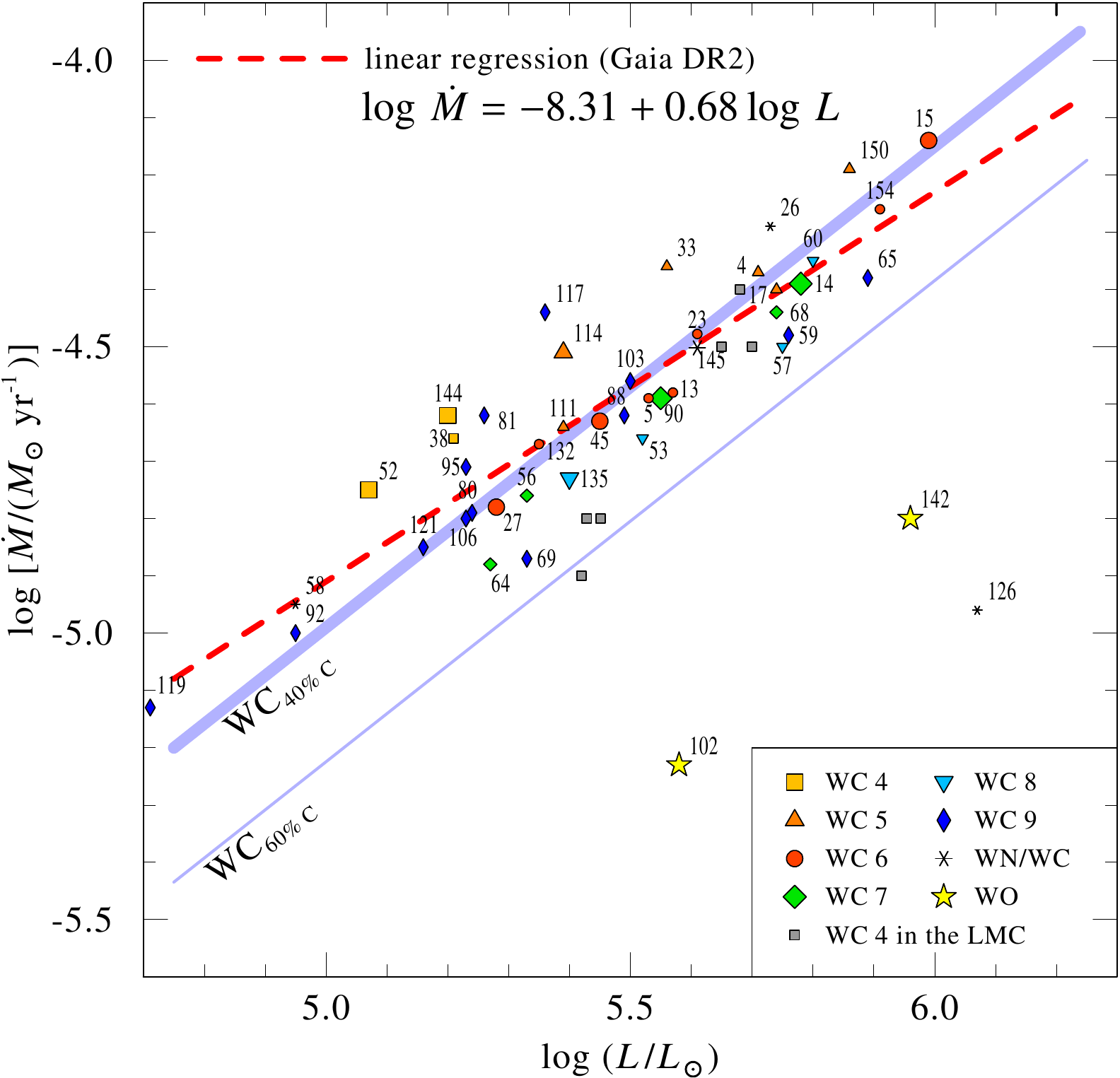}}
  \caption{Empirical mass-loss rates versus updated luminosity for the Galactic WC \& WO stars. 
	  The red-dashed line gives the linear regression fit to these results. The relations proposed by \citet{NL2000}
for WC stars with 40\% and 60\% carbon, respectively, are indicated by the thick and thin shaded lines.
    The results from \citet{Crowther+2002} for the LMC WC4 stars are denoted by gray symbols.}
  \label{fig:mdotlgaia}
\end{figure}

 
  With the revised luminosities being given and adopting the distance-independent parameters ($T_\ast$, $R_\text{t}$, $\varv_\infty$, $D$) from \citet{Sander+2012}, we obtain the revised mass-loss rates via Eq.\,(\ref{eq:mdotcalc}). The results are compiled in Table,\,\ref{tab:wcpar} including the uncertainties as discussed in Sect.\,\ref{subsec:errors}. With now both $L$ and $\dot{M}$ given, we can check how the relation between mass-loss and luminosity is affected and how our results compare to the relation from
  \citet{NL2000}, which is widely applied in stellar evolution calculations. Fig.\,\ref{fig:mdotlgaia} shows the revised results. While the slope of linear regression fit in \citet{Sander+2012} ($0.83$) was in line with \citet{NL2000} obtaining $0.84$, the updated figure suggests that the luminosity-dependence is weaker, yielding
\begin{equation}
  \label{eq:ownmdotl}
  \log \frac{\dot{M}}{M_\odot \text{yr}^{-1}} = -8.31 (\pm 0.29) + 0.68 (\pm 0.05) \cdot \log \frac{L}{L_\odot}, 
\end{equation}
	with the given errors denoting the formal standard deviations obtained by the linear regression. For comparison, we also show the results for the LMC WC stars analyzed by \citet{Crowther+2002}
	in Fig.\,\ref{fig:mdotlgaia}. They tend to pretty much blend into the cloud of Galactic data points, while a closer inspection shows that three of the six objects, namely those with
	slightly lower mass-loss rates (BAT99\,8, BAR99\,9, and BAT99\,90) seem to match the relation of \citet{NL2000} assuming a carbon mass fraction of $0.6$. However, this is a pure 
	coincidence as neither the different iron abundance -- \citet[][]{Crowther+2002} used 0.4 $Z_\odot$ for his LMC models -- is taken into account, nor does the carbon fraction match. Only one of
	them (BAT99\,8) has $X_\text{C} \approx 0.5$ while the other two have a value close to or even below $X_\text{C} = 0.4$. Due to the two groups, the $\dot{M}$-$L$-relation for the LMC WC4 stars is
	much steeper, having a slope of $1.29$.

	The fact that the LMC WC4 stars are at higher luminosities compared to 
	their Galactic counterparts is expected due to the different metallicity $Z$. Nevertheless, it should be noted that this comparison might be slightly biased
	due to the fact that the models used by \citet{Crowther+2002} do not contain ions higher than \ion{Fe}{viii}. Since we see for the Galactic results that especially for early WC subtypes the
	inclusion of high Fe ions is crucial, this might explain why the LMC WC4 stars are reproduced with cooler models. Due to the parameter degeneracy for dense winds, such a cooler ``solution'' 
	in terms of spectral appearance is often easy to find, but unfavorable or even unphysical from a hydrodynamical standpoint as demonstrated by \citet{GH2005}.  
	
	As discussed in length in \citet{Sander+2012}, the finding that most of the models
	align in the $\log R_\text{t}$-$\log T_\ast$-plane, would imply a slope of $0.75$ in the case of the terminal velocity $\varv_\infty$ (and clumping factor $D$) being the same for all stars. Interestingly, the new distances now lead to a flatter slope, while the value in \citet{Sander+2012} was steeper. We can try to eliminate the effect of $\varv_\infty$
	by including an explicit velocity dependence in the relation, leading to
\begin{align}
  \log \frac{\dot{M}}{M_\odot \text{yr}^{-1}} = & -9.47 (\pm 0.30) + 0.62 (\pm 0.04) \cdot \log \frac{L}{L_\odot} \nonumber \\
	 & + 0.44 (\pm 0.08) \cdot \log \frac{\varv_\infty}{\text{km/s}},
\end{align}
 which actually flattens the slope even further.

 To obtain a relation similar to \citet{NL2000} we can also account for the helium abundance, which is slightly different in some of the sample stars. Due to
 $Z = 1 - Y$ for hydrogen-free stars, explicitly accounting for both terms is not useful and we simply obtain
 \begin{align}
   \label{eq:ownmdotlhe}
   \log \frac{\dot{M}}{M_\odot \text{yr}^{-1}}  = & -8.68 (\pm 0.31) + 0.71 (\pm 0.04) \cdot \log \frac{L}{L_\odot} \nonumber \\
  &	- 0.74 (\pm 0.29) \cdot \log X_\text{He}\text{.}
 \end{align}
 While there is indeed a slight steepening in the luminosity dependence, this has to be taken with care as only very few sample stars in \citet{Sander+2012} actually used  
  models with non-grid abundances. Moreover, \citet{Tramper+2016} suggested to introduce a term depending on $X_\text{Fe}$, but since all models in \citet{Sander+2012} use
	the abundance of $X_\text{Fe} = 1.6 \cdot 10^{-3}$, this is not an option inside this sample. While from a theoretical point of view $X_\text{Fe}$ is an excellent parameter
	for a mass-loss relation, as its slope would characterize one of the most important wind driving contributors, the actual determination of $X_\text{Fe}$ in an observation 
	is tough and usually requires excellent UV spectra resolving features of the iron forest which then also has to be precisely reproduced by the atmosphere model. Presently,
	very few attempts for such work exists \citep[e.g.][]{Herald+2001} and the assumption of a ``typical'' iron abundance motivated by the host galaxy is standard,
	not only when analyzing WR stars, but also other hot, massive stars. For WN stars, \citet{Hainich+2015} determined the dependence of $\dot{M}$ on $X_\text{Fe}$ by studying
	the WN populations of different galaxies with different metallicities. This is an excellent approach that holds as long as the spread in $X_\text{Fe}$ inside the WN population 
	of a particular galaxy is small, but as mentioned before this is unfortunately tough to constrain from observations and would require more and better UV spectroscopy, highlighting
	the urgent demand for a new UV telescope filling the oncoming gap after the end of the HST.

	From an observational point of view, assuming a negligible iron abundance spread inside the LMC, a glimpse of a potential $X_\text{Fe}$-dependency of the WC mass-loss rates can be 
	  obtained from comparing our results with those by \citet{Crowther+2002}. Adding further points at other metallicities turns out to be tough. In the SMC, no WC stars are known. 
		There is one WO (SMC AB 8) which is part of a binary system analyzed by \citet{Shenar+2016}. However, mixing WCs and WOs here would be a problem as we know, e.g., from Fig.\,\ref{fig:mdotlgaia},
		that the mass-loss rates of the WO stars are lower than those of WCs with the same luminosity. A comparison of only the WO stars would in principle be an option, but 
		apart from the low number of known objects the question would remain whether a result could also be applicable to the larger WC population.
		
		Instead we use the fact that all known WC stars in the LMC have a similar spectral type (WC4) and compare them to their Galactic counterparts. Two additional data points
		are obtained from \citet{Abbott+2003} and \citet{Abbott2004}, who analyzed several WR stars in different galaxies, including two WC4-5 stars ([MC83] 70 alias MC70, and [MC83] 79 alias MC79) in
		the outskirts of M33, where the metallicity is SMC-like. To eliminate the luminosity-dependence, we use the ``transformed mass-loss rate''
		
		\begin{equation}
		  \label{eq:mdott}
		  \dot{M}_\text{t} = \dot{M} \sqrt{D} \cdot \left( \frac{1000\,\text{km/s}}{\varv_\infty} \right) \left( \frac{L}{10^6 L_\odot} \right)^{3/4}\text{,}
		\end{equation}
 		introduced by \citet{GraefenerVink2013}. The value of $M_\text{t}$ can be understood as the mass-loss rate $\dot{M}$ the star would have, if it had a smooth wind (i.e., $D = 1$), a 
		terminal wind velocity of $1000\,$km/s and -- most importantly here -- a luminosity of $10^6\,L_\odot$. Since the Galactic sample of WC4 stars consists of three stars,
		we also considered the WC5 stars, which essentially have the same $M_\text{t}$ as the WC4 stars. This is a consequence of Eq.\,(5) in \citet{Sander+2012}, leading to all WC4 to WC7 stars 
		in our Galactic sample having $\log \dot{M}_\text{t} \approx -4.06$. For the LMC sample with their more tailored models, the scatter in $\log \dot{M}_\text{t}$ is larger. Nonetheless, 
		for both samples as well as MC70,  the calculation of $M_\text{t}$ is straight forward, but we had to assume $\varv$ for MC79 and choose a typical value for a WC4 of 
		$\varv_\infty = 3000\,$km/s. 
			
\begin{figure}
  \resizebox{\hsize}{!}{\includegraphics{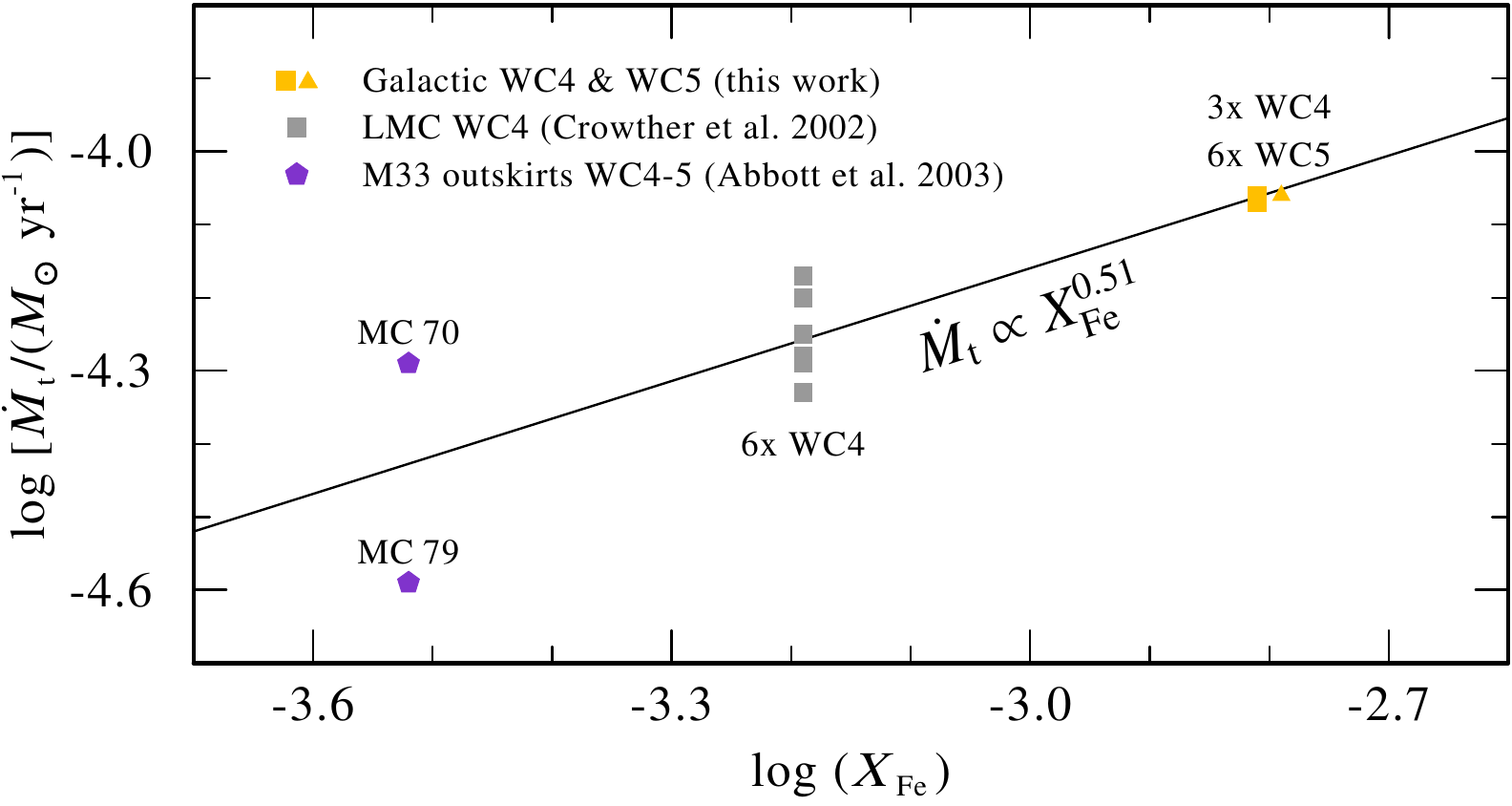}}
  \caption{Transformed mass-loss rate versus Fe mass fraction for WC4 and WC5 stars at different metallicities: The extragalactic results are taken from \citet[LMC]{Crowther+2002}
	   and \citet[M33 outskirts]{Abbott+2003}. The solid line gives the linear regression fit to the data points. The Galactic WC4 and WC5 plot symbols have a slight shift compared to
		the actual datapoint to ensure visibility.}
  \label{fig:mdottfe}
\end{figure}
		
		The resulting distribution of $\dot{M}_\text{t}$ over $X_\text{Fe}$ can be seen in Fig.\,\ref{fig:mdottfe},
		where we also show a linear fit to the double-logarithmic data. Including the two sources in M33, we obtain 
		\begin{equation}
		   \log \dot{M}_\text{t} = 0.51\left(\pm 0.06\right) \cdot \log X_\text{Fe} - 2.62\left(\pm 0.19\right)\text{.}
		\end{equation}
    A dependency of approximately $X_\text{Fe}^{0.5}$ for WC stars is in line with the suggestion from \citet{Crowther+2002}, but significantly more flat than what was found for WN stars by \citet{Hainich+2015}. And even though the data points for M33 in Fig.\,\ref{fig:mdottfe} might have a considerable uncertainty, this result would essentially not change when leaving them out, as we then obtain $X_\text{Fe}^{0.49}$ instead. With Eq.\,(\ref{eq:mdott}) our relation thus points at a dependence for the true mass-loss rate of approximately
		\begin{equation}
		  \dot{M} \propto D^{-1/2}~\varv_\infty~L^{3/4}~X_\text{Fe}^{0.5}\text{.}
		\end{equation}
		From a theoretical point, the issue of metallicity-dependent mass-loss rates for WC stars has been studied by \citet{VdK2005}. While they do not give an explicit $X_\text{Fe}$-dependence,
		their $Z$-definition does not include the self-enriched carbon and oxygen in the WCs. Thus their $\dot{M}(Z)$-slope reflects mainly the change of the primary wind driving contributor, which is
		iron. In total they studied regimes from 10$\,Z_\odot$ down to $10^{-5}\,Z_\odot$ with a slope of $0.66$ in the region. This is slightly steeper than the $0.5$ we obtain here, but 
		given the scatter for the LMC stars and the uncertainty of the M33 data points, still in agreement with what we obtain from the observations.
		
		The well used recipes of \citet{NL2000} do not provide an explicit term for the Fe-dependence, but instead just use a $Z$-term containing all elements beyond He. Of course the two 
		main contributions to $Z$ in WC stars are carbon and oxygen. We made a few test calculations to study the impact of abundance changes on the mass-loss rate using the work ratio
		 \begin{equation}
      Q := \frac{\dot{M} \int \left(  a_\text{rad} - \frac{1}{\rho} \frac{\mathrm{d} P}{\mathrm{d} r} \right) \mathrm{d} r}{\dot{M} \int \left( \varv \frac{\mathrm{d} \varv}{\mathrm{d} r} + \frac{G M_{\ast}}{r^2} \right) \mathrm{d} r }\text{,}
  \end{equation}
        which is a proxy for the global hydrodynamic consistency of the model. In a nutshell, a change in Q between models that are identical in everything but different abundances means that 
				the mass-loss rate should be affected. More information about the concept of the work ratio can for example be found in \citet{GH2005,Sander+2017,Sander+2018}. 
				Our test calculations for a WC5 grid model show that neither a variance of C, nor of O, do really impact the work ratio, at least not if
				the variations are in the observed range of abundances for WC stars, i.e. between $0.2$ and $0.6$ for $X_\text{C}$ and $0.05$ and $0.15$ for $X_\text{O}$. 				
				Of course a proper quantification of the $\dot{M}$-dependence on C and O would be preferable. \citet{VdK2005} did this with a Monte Carlo approach, but a CMF-based calculation with a locally consistent acceleration similar to what has been done for hydrogen-containing WN stars by \citet{GH2008} does not yet exist. However, this is way beyond the scope of the present paper and we thus will come back to this in future studies.
				The concept of the work ratio has also been recently applied to obtain mass-loss rates from grids of models for the bi-stability domain in a study by \citet{Petrov+2016}. In this work
				we only used it as a first step to ensure that the impact is small.
	
				Our finding does not imply that carbon and oxygen are not contributing to the wind driving. However, our test calculations hint that 
				their precise abundances do not really alter the resulting mass-loss rates, at least at Galactic metallicites. Thus, a mass-loss recipe with a simple $Z$-dependence is not properly
				reflecting the physical dependencies of $\dot{M}(X_\text{Fe})$. These recipes -- including the ones from \citet{NL2000} and also our own ones given in Eqs.\,(\ref{eq:ownmdotl}) to (\ref{eq:ownmdotlhe}) -- are best representations to a certain dataset, but their validity is limited to the parameter range they have been established in. Thus, extrapolations of these formulae 
				into different regimes, for example lower masses or different metallicities, are dangerous, and a proper treatment would require a more physically motivated parametrization.

\begin{figure}
  \resizebox{\hsize}{!}{\includegraphics{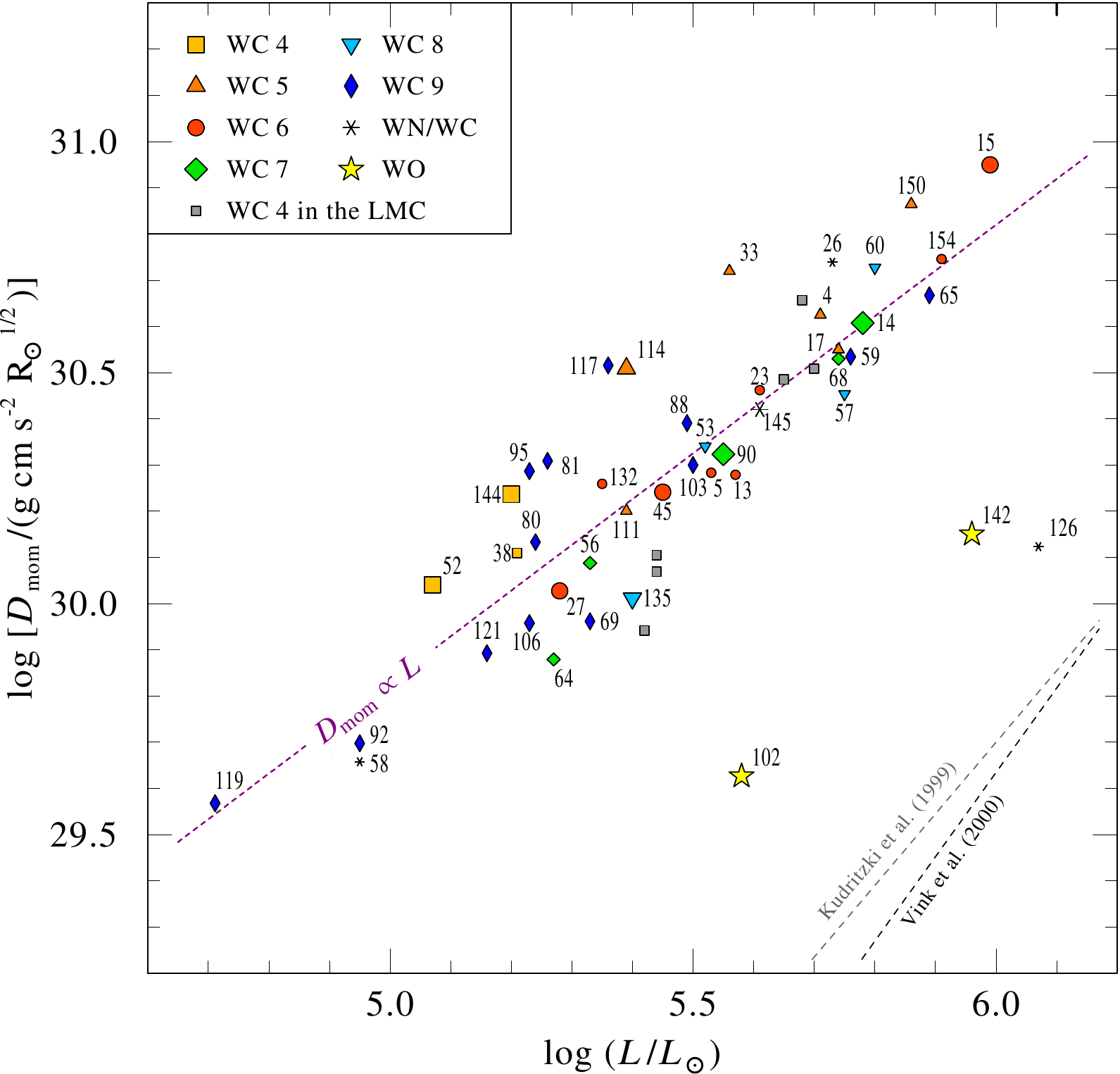}}
  \caption{Modified wind momentum versus updated luminosity for the Galactic WC \& WO stars. 
	  The purple-dashed line gives the linear regression fit to these results. For comparison, the relations for OB stars from \citet{Kudritzki+1999} and \citet{VdKL2000} are
		shown as gray and black dashed lines. The gray symbols without labels denote the results from \citet{Crowther+2002} for the LMC WC4 stars.}
  \label{fig:dmomlgaia}
\end{figure}

  Instead of plotting just the mass-loss rate, we can also show the modified wind momentum $D_\text{mom} = \dot{M} \varv_\infty \sqrt{R_\ast}$ \citep[e.g.][]{Kudritzki+1999} versus
	the luminosity as displayed in Fig.\,\ref{fig:dmomlgaia}. This quantity is motivated by the analytical 
	calculations from \citet{Kudritzki+1989} based on extensions of the CAK \citep*[named after][]{CAK1975} theory for
	line-driven winds, yielding 
	\begin{equation}
	  \dot{M} \varv_\infty \sqrt{R_\ast} \propto \hat{k}^\frac{1}{\alpha-\delta} L^\frac{1}{\alpha-\delta}\text{,}
	\end{equation}
	or simply $D_\text{mom} \propto L^{1/\alpha_\text{eff}}$ with $\alpha_\text{eff} := \alpha - \delta$ and $\hat{k}$, $\alpha$, and $\delta$ denoting the so-called ``force multiplier parameters''. 
	A brief summary of the concept behind this relation is for example given in \citet{Kudritzki+1998}. This means that not only does the 
	modified CAK (mCAK) theory predict a tight relation between $D_\text{mom}$ and $L$, but also that the reciprocal value of the exponent should reflect the force multiplier
	parameter $\alpha$ describing the line strength distribution function. \citet{Kudritzki+1999} found - based on studies of relatively nearby stars - that $\alpha_\text{eff}$
	seems to increase from A-type to early B-type stars and then mildly decreases to a value of $\alpha_\text{eff} \approx 0.65$ for O-type stars. A linear relation fit of
	our revised WC results now yields
	\begin{equation}
	  \log \frac{D_\text{mom}}{\text{g cm s}^{-2}~R_\odot^{1/2}} = 24.88(\pm0.44) + 0.99(\pm0.08) \cdot \log \frac{L}{L_\odot}\text{,}
	\end{equation}
	meaning that essentially $D_\text{mom} \propto L$ and $\alpha_\text{eff} \approx 1$. This would imply a further flattening of the exponent $1/\alpha_\text{eff}$ compared to
	OB stars, something also found for the WN stars in the LMC by \citet{Hainich+2015}. Interestingly, the same exponent of $\approx 1$ was also found for the hydrogen-poor WNL
	stars in M31 \citep{Sander+2014}. These results are in conflict with the numerical calculation of $\alpha_\text{eff} \approx 0$ by \citet{GH2005} for their hydrodynamically
	consistent WC star model. This illustrates that our linear relation between $D_\text{mom}$ and $L$ is only an empirical result and one should be careful when interpreting
	this in the scheme of the mCAK. As for example reviewed in \citet{Puls+2008}, an mCAK parametrization does not hold for optically thick winds with massive line overlap, which is
	essentially the situation we have for Galactic WR winds. The necessity to account for multiline scattering when describing the radiative driving of the wind is also
	reflected in the values obtained for the wind efficiency parameter
	\begin{equation}
	  \eta := \frac{\dot{M} \varv_\infty c}{L}\text{,}
	\end{equation}
	listed in the last column of Table\,\ref{tab:wcpar}. With the exception of the also otherwise odd transition-type star WR 126 (see discussion in Sect.\,\ref{subsec:notable}),
	all objects in our sample significantly exceed the single-scattering limit of $\eta = 1$. For several objects, especially those of early WC subtypes, $\eta$ even exceeds values
	of $10$ or $20$. It is challenging to explain these high values with a purely radiatively driven wind, even when properly accounting for multiple scattering. To shed more light on these
	issue, more hydrodynamically consistent models \citep[e.g.][]{GH2005,GH2008,Sander+2017} covering the dense wind regime are needed.
	
\subsection{Subtype-magnitude relation}
  \label{subsec:mvrel}
	
	With the distances now given, we can check the subtype-magnitude relation used in earlier works. In the past, 
	with only few Galactic Wolf-Rayet stars having a suggested distance, for example from cluster memberships, it 
	was common to assume that the absolute magnitude 	
	\begin{equation}
    M_\varv = \varv - \text{D.M.} - A_\varv
	\end{equation}
  in the narrowband system from \citet{S1968} is approximately the same within one Wolf-Rayet subtype.	
	As long as a few or at least one star with a known distance modulus was available per subtype, this could be used
	to derive the apparent distances for all other WR stars of the same subtype, thus allowing to significantly enlarge
	the sample of studied stars. Of course the underlying assumption
	is rather arbitrary and already the existing scatter, especially seen in the Galactic and M31 WN stars 
	\citep{Hamann+2006,Sander+2014}, clearly questioned this assumption. 
	Now, with the distance moduli based on Gaia DR2, we can simply 
	obtain the absolute magnitudes of all via	
	\begin{equation}
	  M_{\varv,\text{new}} = M_{\varv,\text{old}} - \Delta\text{D.M.}\text{,}
  \end{equation}
	and check how valid the assumption of a roughly constant $M_\varv$ per subtype is.
	
	\begin{figure}
    \resizebox{\hsize}{!}{\includegraphics[angle=-90]{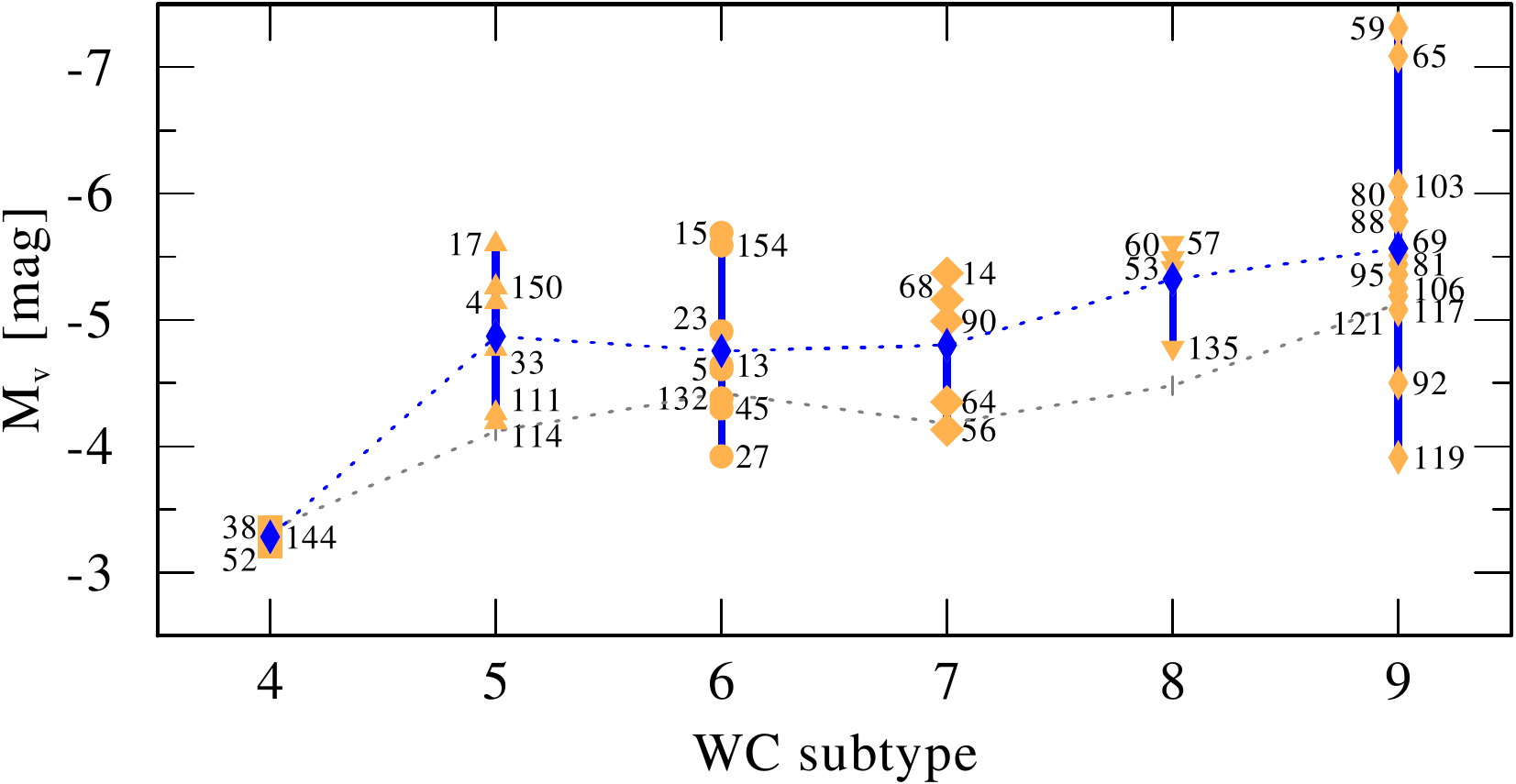}}
    \caption{Mean absolute magnitudes of the WC stars in the Milky Way depicted
by subtype: The orange symbols indicate the positions of the individual stars with the thick blue lines 
marking the span in magnitudes. The blue diamonds connected by the dotted blue line mark the mean value per subtype. 
The gray dotted line denotes the calibration result from \citet{Sander+2012}.}
    \label{fig:galwcmv}
  \end{figure}

\begin{table}
  \caption{Mean absolute magnitudes for the different WC 
    subtypes derived from the revised results based on Gaia DR2 (cf.\ Fig.\,\ref{fig:galwcmv})
		compared to the calibration from \citet[$M^\text{2012}_{\varv}$]{Sander+2012} using stars with distances known from their 
    cluster or association membership}
  \label{tab:wcmvcalib}

  \centering
  \begin{tabular}{lcc}
      \hline
      \hline
Subtype  \rule[0mm]{0mm}{3.4mm} & \multicolumn{1}{c}{$M_{\varv}$ [mag]} & \multicolumn{1}{c}{$M^\text{2012}_{\varv}$ [mag]} \\[0.5mm]
      \hline  
      WC4  \rule[0mm]{0mm}{3mm} & -3.28 &  $-3.34$  \\
      WC5  & -4.87 &  $-4.12$  \\
      WC6  & -4.75 &  $-4.42$  \\
      WC7  & -4.80 &  $-4.18$  \\
      WC8  & -5.32 &  $-4.48$  \\
      WC9  & -5.57 &  $-5.13$  \\
    \hline
  \end{tabular}
\end{table}   
	The so obtained distribution of $M_\varv$ vs. subtype is shown in Fig.\,\ref{fig:galwcmv} and Table\,\ref{tab:wcmvcalib}, where we also denote the
	result from \citet{Sander+2012} in gray for comparison. Only the WC4 stars essentially keep their mean $M_\varv$
	while all other subtypes now have higher mean values. There are two basic reasons for this: First, for several
	subtypes there was just one star with a ``known'' distance in the 2012 sample. This yielded reliable results as long as this distance turns out to be
	confirmed by Gaia, such as for WR144, where the distance inferred from the parallax seems to verify the 
	possible membership of Cyg OB2 claimed by \citet{LS1984} including their distance modulus. However, in other cases
	such as the WC7-star WR68, even the possible membership to the Cir OB1 association did not help, as the distance
	to the association has a significant uncertainty as different studies came to different results in the past. Thus the
	absolute magnitude inferred from the Gaia-based distance now increases by almost $1\,$mag, significantly affecting
	the calibration. 
	Another issue now revealed with the help of the Gaia DR2 distances, is the internal spread in $M_\varv$, putting
	a general question mark behind this method. While for the WC4 subtype the results in Fig.\,\ref{fig:galwcmv}
	seem to hint at a more or less common $M_\varv$, other subtypes show a huge spread in intrinsic magnitudes.
	As is evident in Fig.\,\ref{fig:galwcmv}, this spread is clearly present for all subtypes between WC5 and WC9, with an exception for the WC4 subtype, which
	could be possibly due to low number statistics. This would be in line with the picture that we see for example for the WC8
	subtype: In \citet{Sander+2012}, WR135 was the only WC8 star with a ``known'' distance. While the $M_\varv$ with the distance
	based on Gaia DR2 only changed by $0.04\,$mag, three other WC8 stars are now available, all showing considerably 
	larger values of $M_\varv$. In total, we can confirm the earlier suspicion that an $M_\varv$ calibration can
	at best provide rough estimates for the resulting luminosities when applied to an individual star and thus should
	be handled with great care.

\subsection{Notable objects}
  \label{subsec:notable}	
	
	In this section we comment on specific targets that are especially affected by the revision due to the new
	distance information:
	
	\emph{WR 15:} With a luminosity of $\log L/L_\odot = 5.99$, the WC6 star WR 15 is the most luminous star in our sample, not accounting
	for the strange and slightly suspicious transition type star WR 126 which is discussed below. In contrast to some of
	the other highly luminous WC stars in our sample, it has a rather low relative parallax error of only $9\%$, leading to a robust
	distance of $3.0\pm0.25\,$kpc. WR 15 is considered to be a possible member of a potential concentration of OB stars termed Anon Vel b
	by \citet{LS1984}, for which they also provide a distance of $2.5\,$kpc following \citet{Bassino+1982}. The results based on Gaia DR2 are now 
	more in line with the original suggestion by \citet{Muzzio1979} that WR 15 could be associated with group of slightly more distant OB stars
	at about $3.2\,$kpc.
	
	\emph{WR 59:} WR 59 is a WC9d star and suspected binary candidate due to weak absorption features in the optical spectrum \citep{Williams+2005}. 
	WR 59 underwent the most drastic luminosity revision with now $\log L/L_\odot = 5.76$, a shift by $+0.86\,$dex. The reason
	of the previous underestimation is due to the $M_\varv$-calibration discussed in Sect.\,\ref{subsec:mvrel} that was applied as
	there was no information on the distance earlier. As this star turns out to have a very similar position in the HRD as the CWB WR 65 (see below),
	we mark it with a warning in the results Table\,\ref{tab:wcpar}.
	
	\emph{WR 65:} Another WC9d star is WR 65, which was suspected to by a binary due to optical absorption features by \citet{Williams+2005} 
	 and later classified as CWB by \citet{OH2008} due to significant X-ray emission. In \citet{Sander+2012}, we excluded WR 65 from the $M_\varv$
	 calibration due to the high value of $M_\varv \approx -7$ one would obtain in the case of trusting the supposed association membership to Cir OB1. 
	 The distance based on Gaia DR2 now yields $M_\varv = -7.09$, thus indirectly confirming the membership, but also pointing towards a significant
	 contribution of the companion. In this case, the underlying spectral analysis would not yield the correct stellar parameters, which is why
	 we mark this star in Table\,\ref{tab:wcpar} with a warning, similar to WR 59.
	
	\emph{WR 88:} WR 88 was classified as another WC9 stars until \citet{Williams+2015} revealed WN features in the spectrum. A detailed modeling
	of the atmosphere by the same authors identifies WR 88 as a WC9/WN8 transition-type star with $X_\text{C} = 0.07$, a carbon abundance lower typical for
	WC stars, but already depleted nitrogen ($X_\text{N} = 0.003$) and slightly enriched oxygen ($X_\text{O} = 0.004$). To preserve a homogenous
	sample, we keep the stellar parameters from \citet{Sander+2012} in Table\,\ref{tab:wcpar}, but assign the revised spectral type. Unlike most 
	transition-type stars, the WC9 features in WR 88 are clearly dominant here, so we also keep this object in the bucket of the WC9 stars for the discussions
	and plots, especially as $T_\ast$ stays the same in both studies. 
	
	Rescaling the 
	stellar parameters derived by \citet{Williams+2015} to the new Gaia distance for WR 88 leads to a luminosity of $\log L/L_\odot = 5.33$, slightly lower than
	the value of $5.49$ we obtain using our 2012 model. Consequently, the mass-loss rate is slightly lower as well ($-4.74$ vs. $-4.62$) and the inferred
	stellar mass drops from about $15\,M_\odot$ to $11\,M_\odot$. While these differences are not dramatic and thus do not affect our general conclusions,
	they underline our initial statement that for a detailed study of a particular object more tailored models should be used.

\begin{table}
  \caption{Revised stellar parameters of the single Galactic WO stars based on the atmosphere models used in \citet{Tramper+2015} and the Gaia DR2 based distances from \citet{Bailer-Jones+2018}.
	         The parameters for the two WO2 stars based on the spectral analysis from \citet{Sander+2012} can be found in Table\,\ref{tab:wcpar}.}
  \label{tab:wotramper}
  \renewcommand{\arraystretch}{1.3}
  \centering
  \begin{tabular}{lccc}
      \hline
      \hline
  \rule[0mm]{0mm}{3mm} & \multicolumn{1}{c}{WR 93b} & \multicolumn{1}{c}{WR 102} & \multicolumn{1}{c}{WR 142} \\
      \hline  
      Subtype  \rule[0mm]{0mm}{3mm}              &   WO3  &    WO2  &   WO2  \\
      $T_\ast$\tablefootmark{(a)} [kK]           & $160$  &  $210$  & $200$  \\
      $\varv_\infty$\tablefootmark{(a)} [km/s]   & $5000$ &  $5000$ & $4900$ \\
      $X_\text{He}$\tablefootmark{(a)}           & $0.29$ &  $0.14$ & $0.26$ \\
      $X_\text{C}$\tablefootmark{(a)}            & $0.53$ &  $0.62$ & $0.54$ \\
      $X_\text{O}$\tablefootmark{(a)}            & $0.18$ &  $0.24$ & $0.21$ \\
      $R_\ast$ [$R_\odot$]                       & $0.44$ &  $0.24$ & $0.39$ \\
  $\log L/L_\odot$\tablefootmark{(b)}                          & \multicolumn{1}{r}{$5.04^{+0.17}_{-0.14}$} & \multicolumn{1}{r}{$4.98^{+0.07}_{-0.06}$} & \multicolumn{1}{r}{$5.34^{+0.04}_{-0.04}$}  \\
  $\log \dot{M}/(\text{M}_\odot\text{/yr})$\tablefootmark{(b)} & \multicolumn{1}{r}{$-5.19^{+0.13}_{-0.11}$} & \multicolumn{1}{r}{$-5.27^{+0.05}_{-0.05}$} & \multicolumn{1}{r}{$-4.98^{+0.03}_{-0.03}$}  \\
      $M_\ast$\tablefootmark{(b),(c)} [$\text{M}_\odot$]              &  $8.1^{+1.9}_{-1.2}$  & $7.3^{+0.6}_{-0.5}$ & $11.7^{+0.7}_{-0.6}$  \\[0.5mm]
    \hline
  \end{tabular}
	  \tablefoot{
      \tablefoottext{a}{taken from \citet{Tramper+2015}}
      \tablefoottext{b}{errors based on uncertainties of Gaia DR2 parallaxes converted into distances by \citet{Bailer-Jones+2018}.}
      \tablefoottext{c}{Masses are calculated from luminosity after \citet{Langer89mass} using their $M$-$L$ relation for WC stars.}
  }
\end{table}  
	
\begin{figure}
  \resizebox{\hsize}{!}{\includegraphics{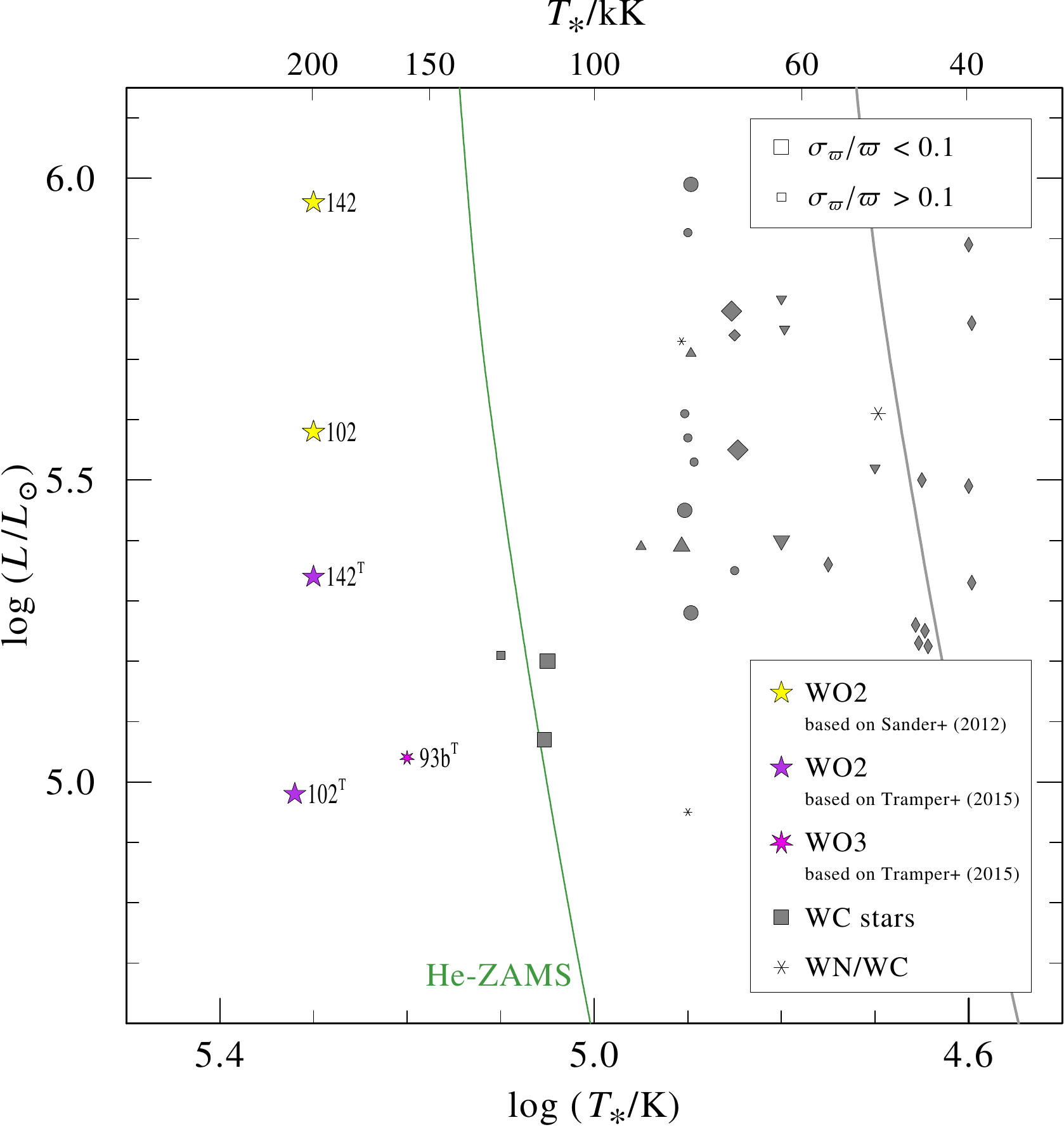}}
  \caption{HRD with highlighted WO star positions: The yellow symbols indicate the positions using the stellar parameters from \citet{Sander+2012}
	 with the revised distances. The purple symbols use the same distances, but apply the stellar parameters from \citet{Tramper+2015}, including
	 the result for the WO3 star WR 93b which was not covered in \citet{Sander+2012}.}
  \label{fig:hrdtramper}
\end{figure}

  \emph{WR 93b:} The WO3 star WR 93b was not analyzed by \citet{Sander+2012} due to the lack of available spectra at that time. Using the 
	atmosphere parameters from \citet{Tramper+2015}, we have calculated the revised radius, luminosity, mass-loss rate, and stellar mass
	based on the new distances from \citet{Bailer-Jones+2018}. The results for WR 93b and the other two single Galactic WO stars are
	listed in Table\,\ref{tab:wotramper} with the HRD positions being shown in Fig.\,\ref{fig:hrdtramper}. As these values are from a different
	set of stellar atmosphere models and the results for the WO2 stars differ greatly between the two approaches as we discuss below, 
	we list them separately and do not include WR 93b in the other diagrams which are solely based on the 2012 sample.
	
	\emph{WR 102 and WR 142:} The WO2 stars WR 102 and WR 142 shared a pretty similar position in the HRD in \citet{Sander+2012}. Now, with the updated
	distances, they mutually differ more in their respective luminosity, with WR 102 being shifted downwards by about $0.1\,$dex and WR 142 shifted upwards by
	more then $0.2\,$dex in luminosity, implying a current mass of about $29\,M_\odot$, making the star a potential progenitor of a massive 
	stellar black hole. \citet{Tramper+2015} analyzed the two WO2 stars with the help of new spectra and
	derived more fine-tuned abundances, yielding quite similar stellar temperatures, but different luminosities of $\log L/L_\odot = 5.45$
	for WR 102 and $5.39$ for WR 142. One would expect that this is mostly due to the different distances assumed for the objects at that time,
	but rescaling their results to the new distances based on Gaia DR2 would now lead to $\log L/L_\odot = 4.98$
	for WR 102 and $5.34$ for WR 142. For both stars, this is about $0.6\,$dex lower than the revised results based on our models
	which we illustrate in Fig.\,\ref{fig:hrdtramper}. 
	
	Such a huge discrepancy is stunning and raises the question for their origin. Are the different  
	stellar atmosphere models responsible, of are there other issues, for example in the proper reconstruction of the SED, which
  differs also in terms of technique between \citet{Sander+2012} and \citet{Tramper+2015}. In \citet{Sander+2012}, we used additional photometry
	(Smith $b$ and $\varv$ in the optical, JHK in the near-IR) to constrain the SED, while \citet{Tramper+2015} derived a Johnson V-band magnitude
	from their spectral observation that was then compared to the literature to obtain a flux calibration.
	We carefully revisited our spectral analyses for both stars and could not find any issues in our SED reconstruction. Unfortunately, the 
	models in \citet{Tramper+2015} do not account for Fe ions higher than \ion{Fe}{x}, which play a significant role in the temperature regime 
	of these stars, but whether this can account for the huge discrepancy remains unclear.
	
	The lower luminosities resulting from the updated distances applied to the models from \citet{Tramper+2015} interestingly
	have only a moderate effect on the inferred mass-loss rates with about $0.2\,$dex difference for WR 142 and less than $0.1\,$dex 
	difference for WR 102. On the other hand, the consequences for the stellar masses are dramatic as they sensitively depend on the
	luminosities. A comparison of the masses for WO2 stars between Table\,\ref{tab:wcpar} and \ref{tab:wotramper} reveal that those
	inferred from our models are more than a factor of two higher. As we do not see an issue with our spectral analysis from \citet{Sander+2012} 
	apart from the fine-tuning of the abundances, we use the revised parameters from Table\,\ref{tab:wcpar} for the discussions
	throughout the rest of this work.
	
	WR 142 was the first WO star to be detected in X-rays with faint, but hard spectrum \citep{Oskinova+2009,Sokal+2010}. X-rays in single WC stars are not measured,
	but neither are there spectral hints for binarity, nor is the inferred $L_\text{X} \approx 1.3 \cdot 10^{31}\,$erg/s high enough to make WR 142 a
	candidate for a CWB. As WO winds are weaker compared to WC winds from stars with a similar luminosity (cf. Fig.\,\ref{fig:mdotlgaia}),
	this might explain why X-rays can be measured for a putatively single star here while they cannot for the WC single stars with more dense winds.
	
	\emph{WR 119:} With $\log L/L_\odot = 4.7$, the WC9d star WR 119 is now the least luminous star in our sample, being shifted down 
	by $0.5\,$dex compared to our 2012 paper. There is considerable uncertainty of about $25\%$ in the underlying parallax measurement,
	but even if the luminosity would be higher by about $0.3\,$dex, WR 119 would still be among the few WC stars marking the bottom end
  of the luminosity range. The star was part of the recent variability study by \citet{Desforges+2017}, but did not show any particularly
	special behavior compared to other WC9d stars.
	
	\emph{WR 126:} For WR 126, the luminosity was significantly revised from $\log L/L_\odot = 5.43$ to $\log L/L_\odot = 6.07$, making it the
	most luminous object in our sample. It is a somewhat odd outlier classified as WC5/WN, showing a spectral appearance with
	dominant WC features, but also prominent nitrogen lines, which is very rare as other WC stars do not show any trace of
	nitrogen while most transition-type stars have dominating WN features and are thus classified as WN/WC instead of WC/WN.
	While there is so far no clear evidence for binarity, the high luminosity combined with the relatively weak emission lines are suspicious.
	The WC/WN star might be part of a binary or multiple system, which would lead to a dilution of emission lines and an
	overestimation of the luminosity of the WR component. Given the large distance between $7.6$ and $11.6\,$kpc according
	to the Bayesian method from \citet{Bailer-Jones+2018}, about 2--3 times more than previously assumed, 
	WR 126 might even be a small, unresolved cluster harboring a WC as well as a WN star instead of a transition-type object.
	
	\emph{WR 145:} WR 145 is a WR+O binary system with a WN7/WCE transition-type primary. Using optical spectra, \citet{Muntean+2009} 
	derived a solution for the system with $M_\text{WR} = 18\,M_\odot$ and $M_\text{O} = 31\,M_\odot$. Applying the technique from \citet{Demers+2002},
	they splitted the spectral components and determined the companion to be an0 O7 V((f)). Moreover, the system is the brightest WR X-ray 
	source in Cyg OB2 and was recently studied by \citet{Rauw+2015}, who deduced that the bulk of the X-ray emission might stem from the O-star
	companion while only a smaller amount originates in the wind-wind collision zone. A rough estimate based on the new Gaia distance gives $L_\text{X} \leq 10^{33}\,$erg/s.
	
	\emph{WR 150:} The WC5 star WR 150 is the sample star with the smallest measured parallax of only $\varpi = 0.024\,$mas. Its error 
	is $\sigma_\varpi = 0.025\,$mas, making it the only star in our sample which has a relative error that is as large as the
	measurement itself. A simple inversion of the parallax would place the star at a distance of more than $40\,$kpc, way
	outside the Milky Way. The Bayesian method by \citet{Bailer-Jones+2018} gives a distance between $8.3$ and $11.9\,$kpc,
	which seems to be more likely as it would place WR 150 more or less in an outer spiral arm (cf. Fig.\,\ref{fig:galwcpos}) and leads to a reasonable
	luminosity of $\log L/L_\odot = 5.86$. 
	
\section{Discussion of the evolutionary status}
  \label{sec:evol}  

\begin{figure}
  \resizebox{\hsize}{!}{\includegraphics{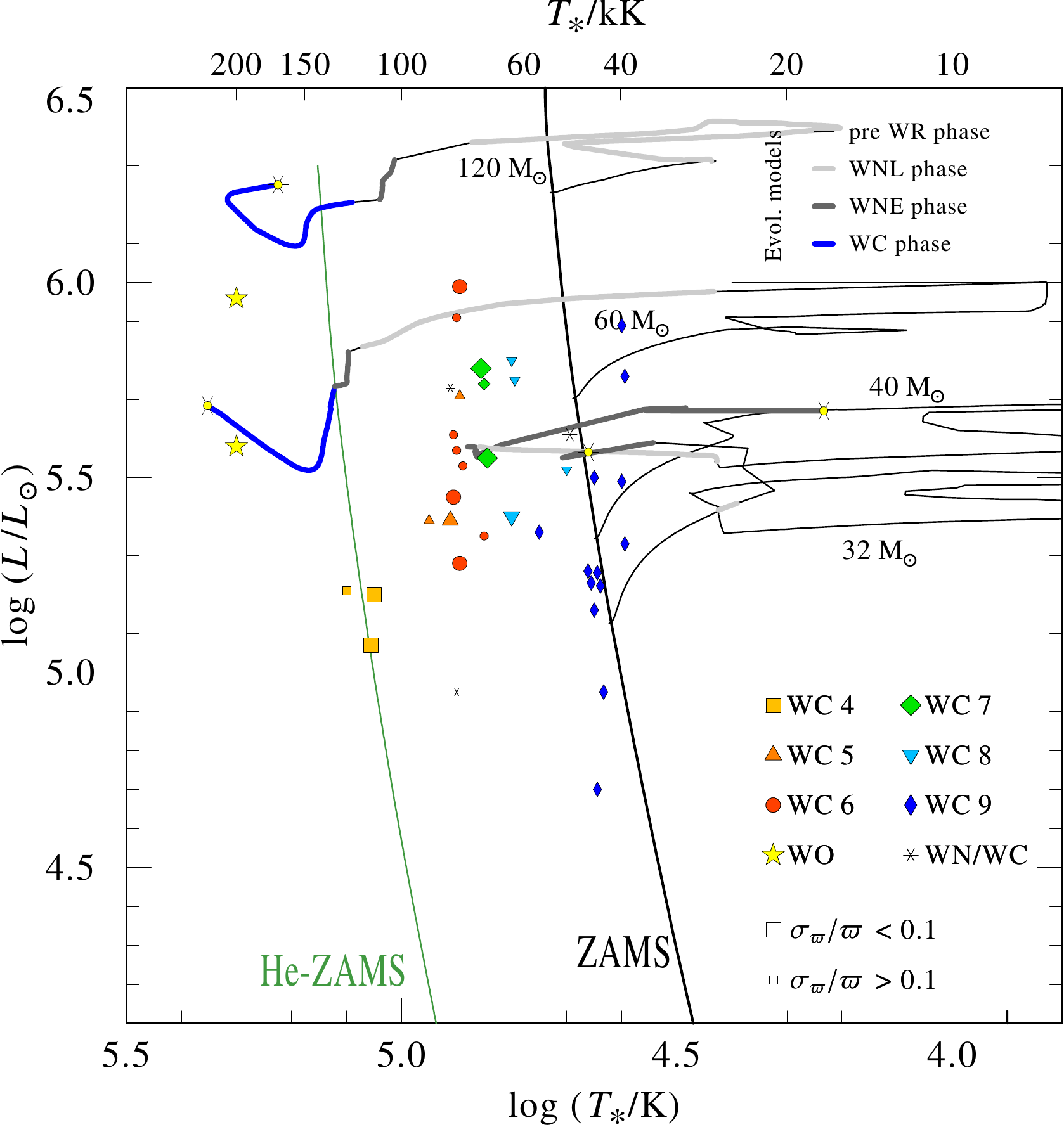}}
  \caption{HRD with the WC \& WO star positions compared to the evolutionary tracks
	 from \citet{Ekstroem+2012} without initial rotation.
The thick lines indicate the WR phases of the tracks.}
  \label{fig:hrdgenfnorot}
\end{figure}

\begin{figure}
  \resizebox{\hsize}{!}{\includegraphics{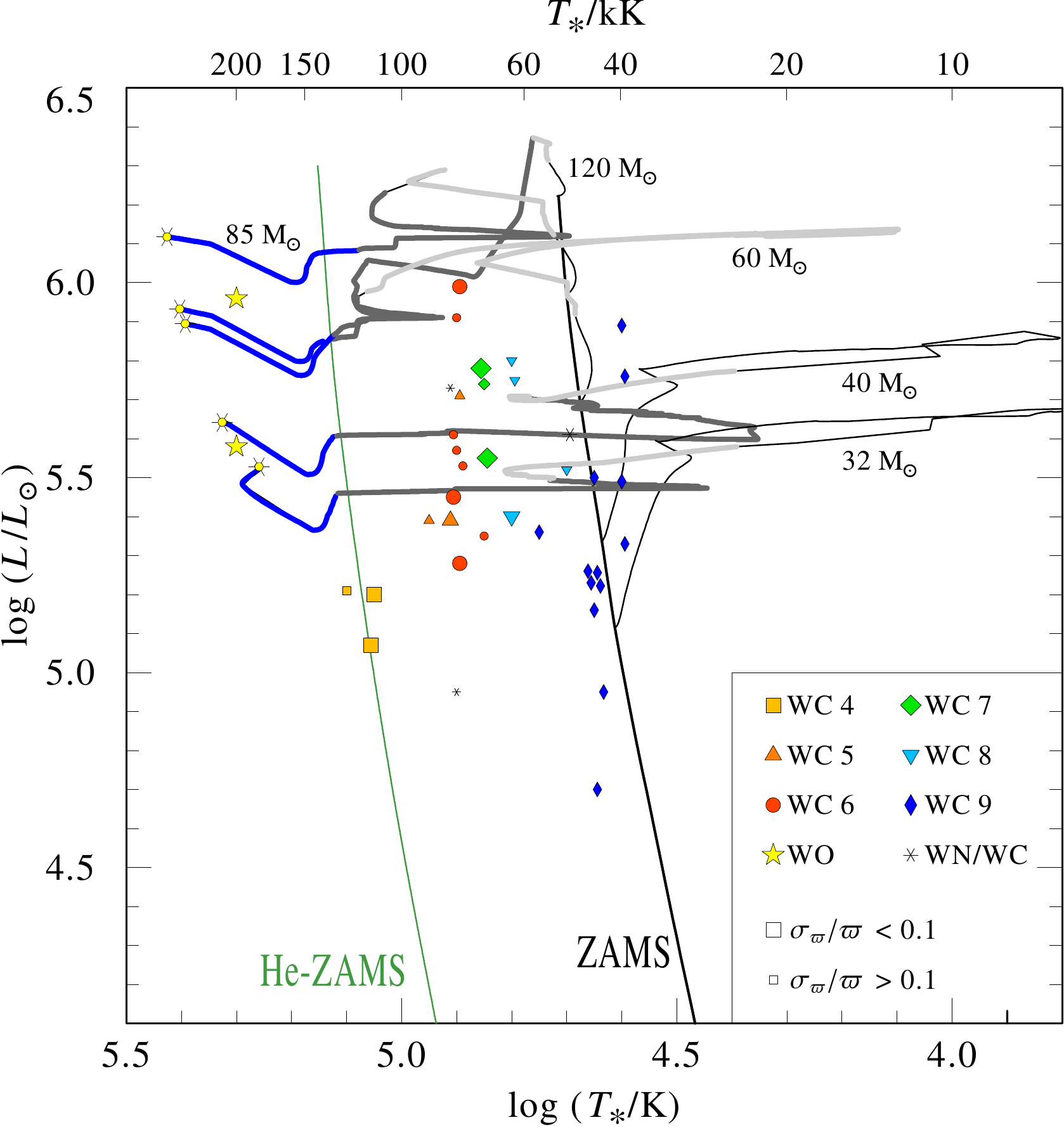}}
  \caption{HRD with the WC \& WO star positions compared to the evolutionary tracks
	 from \citet{Ekstroem+2012} with an initial rotational velocity of 0.4 $\upsilon_\text{crit}$.
The thick lines indicate the WR phases of the tracks.}
  \label{fig:hrdgenfrot}
\end{figure}

\begin{figure}
  \resizebox{\hsize}{!}{\includegraphics{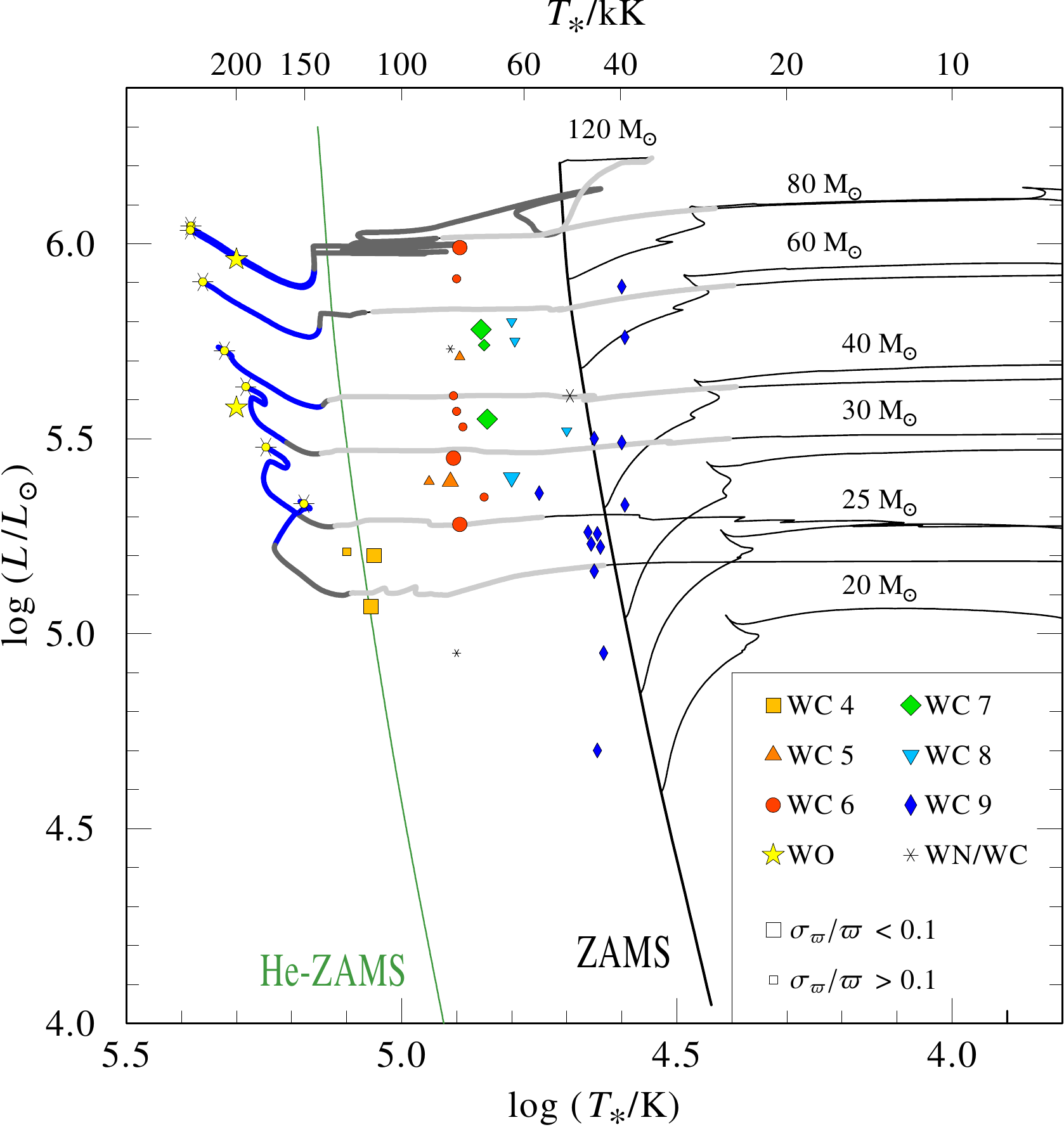}}
  \caption{HRD with the WC \& WO star positions compared to the evolutionary tracks
	 from \citet{CL2013} with an initial equatorial rotational velocity of $\upsilon_\text{rot,ini} = 300\,$km/s.
The thick lines indicate the WR phases of the tracks.}
  \label{fig:hrdfranecrot}
\end{figure}
  
\subsection{Comparison with evolutionary tracks}
  \label{subsec:tracks}

  In order to form a classical, helium-burning Wolf-Rayet star, a star needs to lose its outer, hydrogen-rich layers.
	This can be done essentially in two ways: One possibility is that the star peels off its outer layers via intrinsic
	mass loss. The latter is believed to originate in either stellar winds \citep[``Conti scenario'', named after][]{Conti1976}, or by episodic, massive outbursts
	due to instabilities \citep[e.g.][]{Langer+1994}, or continuum-driven eruptions \citep[e.g.][]{SmithOwocki2006}. Depending
	on the initial mass, both mechanisms might play a role during the lifetime of a massive star. Alternatively, a WR star can be 
	produced in a binary system where Roche lobe overflow (RLOF) can be an important mass loss channel for the donor star \citep[e.g.][]{Paczynski1967}.
	However, so far, only a few WR stars have been convincingly suggested to have formed via binary interaction \citep[e.g.][]{Groh+2008,Shenar+2016}.
	
	In Figs.\,\ref{fig:hrdgenfnorot} and \ref{fig:hrdgenfrot}, we compare the revised positions of the Galactic WC
	stars in the HRD with the Geneva single-star evolution models for $Z = 0.014$ from \citet{Ekstroem+2012} without and with
	initial rotation, respectively. The evolutionary tracks are color-coded to highlight WR stages, following the usual
	assumption that these could be identified from surface compositions. A star with $T_\ast > 25\,$kK is considered a WR 
	star if the hydrogen abundances is $X_\text{H} \leq 0.4$. It is marked as a hydrogen-rich WN until $X_\text{H} \leq 0.05$.
	Below this abundance, we mark the WR as a hydrogen-free WN, or as WC/WO if $X_\text{C} > 0.2$. While the precise limits for the abundances
	and temperature ranges differ, these	``buckets'' are essentially corresponding to what has been labeled as WNL and WNE stars in an
	evolutionary sense \citep[cf. e.g.][]{Hamann+2006,Sander+2012,Georgy+2012}. However, since the terms ``late'' (WNL) and ``early'' (WNE)
	originally refer to the spectral type and the study of WR populations at lower metallicities have shown that the finding that early-type
	WN stars are hydrogen-free does not hold there \citep[e.g.][]{Hainich+2014,Hainich+2015,Shenar+2016}, we refrain from 
	over-stressing this potentially confusing terminology. 
	
	Assuming single-star evolution, the models with initial rotation are much better in reaching the area
	of the HRD covered by the WC stars than their counterparts without rotational mixing, which could explain only the
	objects with the highest luminosities. Nevertheless, also the models accounting for rotation do not
	reach $\log L/L_\odot < 5.3$. While several stars have higher luminosities after the current revision, various
	ones remain in this ``lower'' luminosity regime, in particular all studied
	WC4 and some of the WC9 stars. Thus the problem
	already discussed by \citet{Georgy+2012} for the original 2012 results remains, namely that the low luminosity WC stars would require higher mass-loss during
	their previous evolutionary stages. The Geneva models by \citet{Ekstroem+2012} use the formula by \citet{NL2000}
	for all WR stages except for the WNh stage (also termed WNL stage as these mostly co-incides with late WN subtypes
	for solar metallicity), where the recipe from \citet{GH2008} based on a set of hydrodynamically-consistent models 
	is implemented. As denoted by \citet{Georgy+2012}, these theoretically-driven approach leads to lower mass-loss rates
	in the WNh stage than usually inferred by empirical studies of WR stars \citep[e.g.][]{Hamann+2006,Liermann+2010}.   
	
 	Using the Bonn Evolution Code (BEC), \citet{Yoon+2017} argues that mass-loss rates slightly increased by a factor
	of $1.58$ compared to \citet{NL2000} would be enough to explain WC and WO stars found with $\log L/L_\odot < 5.3$. 
	Such a raise in the mass-loss rate could be achieved by assuming a lower clumping of $D = 4$ instead
	of $D = 10$. This would raise questions regarding a sufficient radiative wind driving. 
	Atmosphere models where the hydrodynamic equation of motion is consistently included, in particular the WC results from \citet{GH2005}, but also 
	results for other types of hot stars \citep[e.g.][]{GH2008,Sander+2017,KK2017}, point rather towards $D \geq 10$.
	
	To study the how sensitive our conclusions are to the different choice of stellar evolution models, we 
	compare our results in Fig.\,\ref{fig:hrdfranecrot} to the tracks accouting for rotation from \citet{CL2013} obtained with the 
	Frascati Raphson Newton Evolutionary Code (FRANEC).
  Their evolutionary models are calculated for the same initial metallicity of $Z = 0.014$ as the Geneva models. Nevertheless,
	the FRANEC models differ in several details from the Geneva models, in particular the treatment of mixing including the assumptions
	for the initial rotation which is set to a fixed initial value of $\upsilon_\text{rot,ini} = 300\,$km/s here compared to $0.4\,\upsilon_\text{crit}$
	in the Geneva models. While there are significant differences in terms of the shape of the tracks between Figs.\,\ref{fig:hrdgenfrot} and \ref{fig:hrdfranecrot},
	the most noticeable result of the FRANEC models is probably the lower initial mass of only $20\,M_\odot$ required to reach the WR stage. However,
	one should keep in mind that none of the models up to $M_\text{ini} = 30\,M_\odot$ reaches our typical WC carbon mass fraction of $X_\text{C} = 0.4$ but instead
	end with $X_\text{C} \approx 0.17$ (cf.\,Sect.\,\ref{subsec:errors}). Thus, the WC phase of the tracks indicated in Fig.\,\ref{fig:hrdfranecrot} is defined as
	$X_\text{C} > 0.15$ which should still provide a WC-type spectrum.
	
	Despite the lower luminosities reached by the FRANEC models, the low carbon abundances of their lower mass tracks reveal that
	they essentially show the same discrepancies as the Geneva tracks. Not accounting for the temperature difference as this might
	be due to inflation which we discuss below, they especially cannot reproduce the evolved status of the WC stars with $\log L/L_\odot < 5.4$ as
	their carbon surface mass fractions are higher. A particular noticeable shortcome here are the WC4 stars with $X_\text{C} \geq 0.4$. Thus the FRANEC
	models give some hints how the WR stage could be reached by initial masses as low as $20\,M_\odot$, that is just above the observed upper limit of type II-L and II-P 
	supernova (SN) progenitors. However, similar to the Geneva models they do not provide a viable scenario where enough mass is
	removed or processed such that the observed lower-luminosity WC stars is reproduced.		
	
	Other ways to remove the necessary amount of mass without invoking a binary scenario would be episodic mass-loss
	during a luminous blue variable (LBV) phase or higher mass-loss during the red supergiant (RSG) stage. Especially in the ``lower'' luminosity range,
	the stars might pass through the RSG stage, and calculations by \citet{Vanbeveren+1998,Vanbeveren+2007} have demonstrated
	that higher $\dot{M}$ could bring massive stars back to the blue side of the HRD. However, calculations by \citet{Meynet+2015}
	revealed that higher mass-loss rates during the RSG stage essentially just lower the time of a star spent as a
	red supergiant, but do not increase the total amount of mass lost during this stage considerably. Thus, a higher 
	$\dot{M}$ during the RSG stage would have little impact on the obtained (synthetic) WR population.
	LBVs could help to remove the necessary amount of mass as they appear at a relative wide
	range of luminosities \citep[e.g.][]{Humphreys+2016,Smith+2018}. However, there is an ongoing debate whether LBVs
	are isolated as claimed by \citet{SmithTombleson2015} or associated with O-stars and RSGs as \citet{Humphreys+2016} deduced based 
	on a sample of LBVs in M31 and M33. \citet{Smith+2018} have pointed out that, based on the new Gaia DR2 data, well 
	known Galactic LBVs and LBV candidates with a WR-type spectrum at quiescence such as WR 31a (Hen 3-519) and WR 31b (AG Car) 
	turn out to be at large distances and could thus not be associated with the O-stars projected near to them on the sky in the Car OB association.
	
	\begin{figure}
  \resizebox{\hsize}{!}{\includegraphics{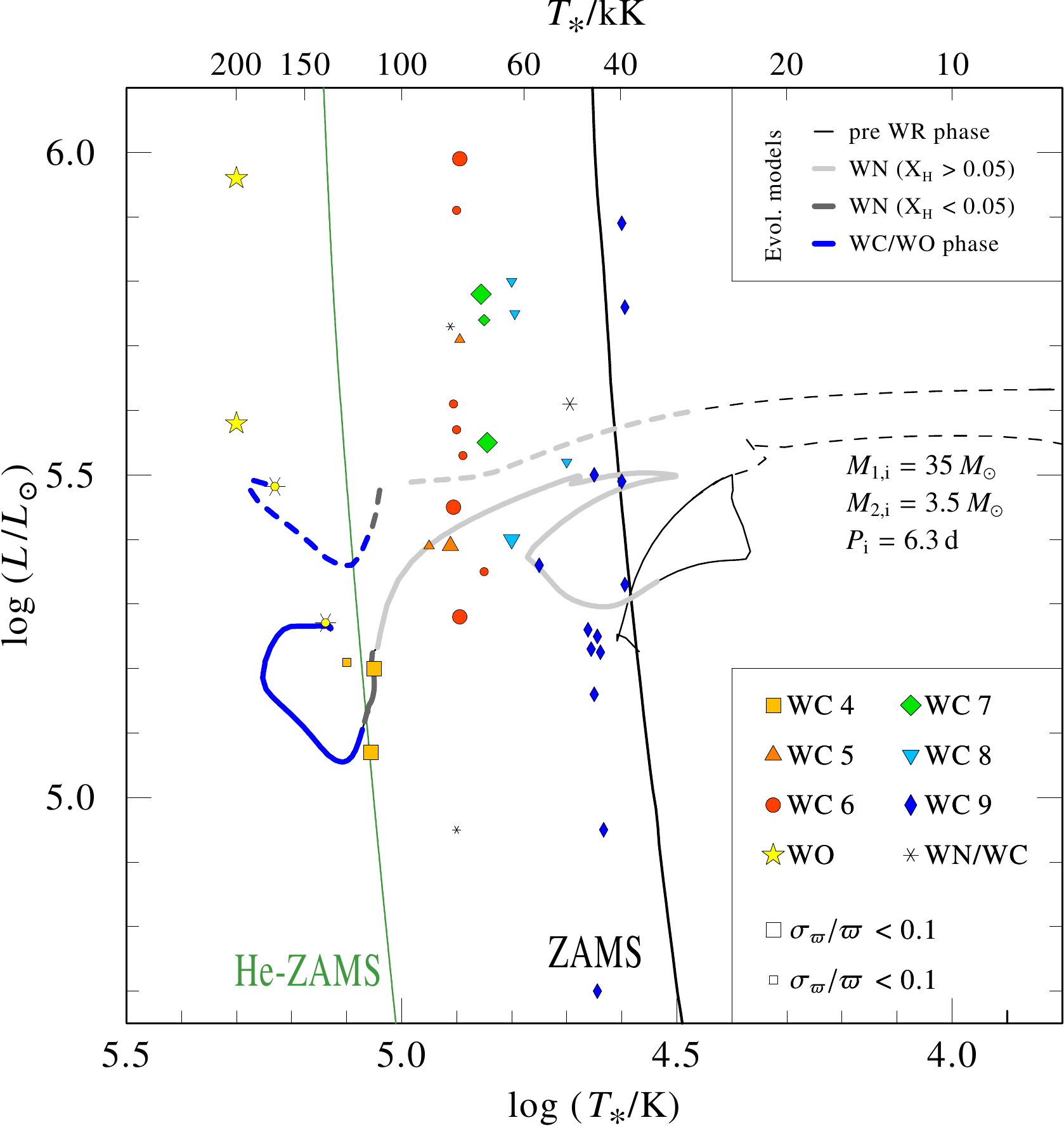}}
  \caption{HRD with the WC \& WO star positions compared to a BPASS binary evolution track. Only the evolution of the
	         primary is shown, as the secondary has a way lower mass and remains on the main sequence at $\log L/L_\odot \approx 2.2$
					 during the whole lifetime of the primary.
					 For comparison, also the corresponding single star evolutionary track for the primary is shown as a dashed line.
           The different colors indicate the different WR phases, defined via surface abundances.}
  \label{fig:hrdbinexample}
\end{figure}
	
	As mentioned in the beginning of this section, binary evolution is an alternative channel to remove enough mass to reach the 
	lower luminosities from our sample, as demonstrated for example by the low masses for type Ibc SN progenitors
	reached in the calculations by \citet{Yoon+2010}. While a detailed study for particular objects is beyond the scope of this paper, we illustrate
	in Fig.\,\ref{fig:hrdbinexample} that the positions of the WC4 stars can be reached with a binary evolution scenario.
	For this example we employ evolutionary tracks calculated with version 2.0 of the Binary Population and Spectral Synthesis
	\citep[BPASS, e.g.][]{ES2009,Eldridge+2017} code. In addition to the physics involved in single-star evolution, BPASS also
	accounts for binary mass transfer. The models do not account for rotationally induced mixing or tidal interactions, but
	this is not necessary for this purpose. In our example, a binary system with a primary of $35\,M_\odot$ and a secondary
	of $3.5\,M_\odot$ (i.e., a mass-ratio of $q = 0.1$) is shown, where the primary transfers mass onto the secondary when
	expanding after the end of its main sequence (MS) phase. Eventually this leads to the effect that the WR stage is reached
	at lower luminosities compared single star evolution, which is also
	depicted in Fig.\,\ref{fig:hrdbinexample}.
	
	Despite the success for the WC4 stars, the problem for the later subtypes, in particular the WC9 stars, remains even when invoking
	binary evolution. While scenarios can be found that cross this area of the HRD as WR stars, none of them has already reached the
	WC stage. As discussed in Sect.\,\ref{subsec:lumo}, the effective temperatures of stellar evolution models are based on
	a simplified treatment of the stellar atmosphere \citep[cf.][]{Groh+2014} and thus the temperature discrepancy could be seen as kind of an ``outer boundary problem'' of the stellar
	evolution models as the later WC subtypes might have inflated envelopes. In particular, according to 
	the results of \citet{Graefener+2012}, the winds of late WC stars with lower luminosity would need to be clumped with $D \geq 10$ for
	reaching the lower temperatures due to inflation, which is on the order of what is derived from electron scattering wings.
	Partially based on the WC results from \citet{Sander+2012} and applying an approximate method to estimate the optical depth
	of the sonic point \citep[cf.][]{VinkGraefener2012}, \citet{GraefenerVink2013} calculated a set of optically thick wind models and
	discovered that the WC and WO stars obey a relation of $P_\text{rad}/P_\text{gas} \approx 80$ at the sonic point. If a star is near the Eddington limit,
	a smooth connection between the outer stellar envelope and their wind models could only be achieved in two cases: In the first one, the sonic point is
	located below at hot Fe opacity peak, which would lead to a compact solution like we get for the WO and early WC subtypes. In the second
	one, there is an extended atmosphere, where the sub-photospheric layers at the location of the sonic point, which is then at lower temperatures above the 
	hot Fe opacity peak, are clumped. \citet{GraefenerVink2013} further argue that due to the concordance of their results with \citet{Glatzel2008}, 
	strange-mode instabilities in the inflated layers might be the origin of WR wind clumping as such.	
	The idea of extended or inflated envelopes for later WC subtypes is further backed by calculations for thick WR winds 
	from \citet{Graefener+2017} who found no smooth wind solutions in this luminosity regime.
	In addition, \citet{ME2016} 
	pointed out, on the basis of pure helium star evolution, that to reach the parameters of the WC9 stars with evolutionary models, 
	inflation alone would not be sufficient as their model counterparts would still have nitrogen which is not observed. They suggest 
	that additional mixing, for example by rotation,
	could help to process the nitrogen into different elements, thus removing it from the surface.

  In total, we can conclude that while the status of the WC stars as evolved, core-helium burning object is beyond doubts, the details of the
	evolutionary path toward this stage still remain unclear to a significant degree. While a certain fraction even of the nowadays
	single WC stars might have a binary history, and some of them might have a yet undetected, low-luminosity companion, the evolutionary
	models from different groups can in principle provide single-star scenarios for WC stars with $\log L/L_\odot \gtrsim 5.4$, thus already
	covering about half of the studied sample. They do not reproduce the temperatures of the late WC subtypes, but this is also an
	issue for binary evolution scenarios. Adjustments of the mass-loss rates in certain evolutionary stages might improve the coverage
	of the HRD regions. Thus, we conclude that single-star evolution is likely not the only, but still a viable channel to obtain WC and WO 
	stars at Galactic metallicity.
	
	\subsection{Evolutionary scenarios for different mass regimes}
  \label{subsec:scenario}
	
	Breaking single-star evolution down into different mass regimes, we have to note that there seems to be still an upper luminosity limit
	for WC stars, but now at $\log L/L_\odot \approx 6.0$ instead of the previously $5.6$ deduced in \citet{Sander+2012}. Thus, for the most massive
	stars with an initial mass $M_\text{i}$ of more than $60$ to $80\,M_\odot$, the scenario inferred by \citet{SmithConti2008} and discussed
	in our 2012 paper is the prime 
	candidate. Introducing the WNH notation (with captial H for distinction) from \citet{SmithConti2008} for luminous, hydrogen-rich 
	WN stars, which are seen as core-hydrogen burning, pre-LBV stars, the evolutionary sequence in the uppermost mass regime reads
	\begin{equation}
	  \text{O} \rightarrow \text{Of/WNH} \leftrightarrow \text{LBV} \left[ \rightarrow \text{WNL}_\text{H-reduced} \right] \rightarrow \text{SN\,IIn}\text{.}
	\end{equation}
	The bracket term suggests that a late-type WN stage with reduced, but not necessarily zero hydrogen could follow the LBV stage
	before the eventual SN explosion. Such stars with low hydrogen mass fractions are found
	among the WNL stars and with parameters close to the so-called ``S Dor instability strip'' \citep[see, e.g.][]{Sander+2014}
	where various known LBVs are located during quiescence. Further evidence for such a scenario are the recent analysis of
	the progenitor spectrum of SN2015bh \citep{BG2018} revealing typical parameters of an LBV candidate	and also the observation of
	SN2014C which turned from a type Ib into a type IIn SN during approximately one year, revealing the presence of a massive shell of
	hydrogen-rich material likely from an earlier, but still relatively recent outburst \citep[][]{Margutti+2017}.

  For lower masses where $M_\text{i}$ is still above the upper mass limit for RSGs, the revised luminosities now lead to the conclusion
	that the WC stage is probably not skipped as originally speculated in \citet{Sander+2012}, thus leading to a sequence like
	\begin{equation}
	  \text{O} \rightarrow \text{Of/WNH} \leftrightarrow \text{LBV} \rightarrow \text{WN} \rightarrow \text{WC} \rightarrow \text{WO} \rightarrow \text{CC}
	\end{equation}
	with CC denoting core collapse. This collapse might come with a SN explosion, but there is also increasing evidence
	from both theoretical and observational side during the last years \citep[e.g.][]{Smartt2015,Sukhbold+2016,Adams+2017} that
	a good fraction of massive stars might fail to explode as a SN but collapse quietly into a compact object, which in our mass
	regime would most likely be a black hole. In case of an exploding WO star, the corresponding supernova would have to be
	of type Ib or Ic, depending on the amount of helium left at the time of collapse. Nevertheless, SN explosion models that
	try to reproduce the observed properties of such a SN with a Wolf-Rayet progenitor turn out to be difficult and highly
  asymmetric explosion scenarios need to be invoked \citep[e.g.][]{Dessart+2017}.
	
	\begin{table}
  \caption{Absolute Johnson magnitudes for selected WC and WO stars scaled to the same luminosity of $\log L/L_\odot = 5.5$. Corresponding magnitudes 
	         for different luminosities can be obtained by adding or subtracting one magnitude per $0.4\,$dex shift in luminosity.}
  \label{tab:magexamples}
  \centering
  \begin{tabular}{lcccc}
      \hline
      \hline
  \rule[0mm]{0mm}{3mm} & \multicolumn{1}{c}{WR 102} & \multicolumn{1}{c}{WR 52} & \multicolumn{1}{c}{WR 13} & \multicolumn{1}{c}{WR 103} \\
      \hline  
      Subtype  \rule[0mm]{0mm}{3mm}               &  WO2   &   WC4   &    WC6   &   WC9  \\
      $M_\text{U}$                                & $-3.0$ & $-5.4$  &  $-5.7$  & $-7.2$ \\
      $M_\text{B}$                                & $-1.8$ & $-4.2$  &  $-4.5$  & $-6.2$ \\
      $M_\text{V}$                                & $-1.7$ & $-4.0$  &  $-4.3$  & $-6.1$ \\
    \hline
  \end{tabular}
\end{table}  
	
	The idea, that the bulk of massive stars with $M_\text{i} \ga 18\,M_\odot$ might collapse silently, or at least without a bright SN outburst, is observationally
	supported from two sides: first, apart from debates about type IIn SNe, no type II SN progenitor with $\log L/L_\odot > 5.1$ has been found \citep{Smartt2015}. Secondly,
	there is the lack of known progenitors for type Ibc SNe with only one possible progenitor candidate discovered very recently \citep{VanDyk+2018}. This has lead to the idea
	that the bulk if not all type Ibc SNe could stem from progenitors with $M_\text{i} = 8$--$20\,M_\odot$ in a binary system \citep{Bersten+2014,Eldridge+2015,Smartt2015}.
	However, an upper mass cut for SNe at about $18\,M_\odot$ would be challenging with regards to our current picture of nucleosynthesis \citep{BW2013}, enforcing fine-tuning
	on some nuclear reactions and also constraints on wind mass-loss rates to avoid a nonobserved overabundance of carbon. 
	
	The nondetection of type Ibc SNe progenitors has lead to magnitude limits that have been compared to potential progenitors. For the type Ic SN 2004gt,
	\citet{GalYam+2005} derived $M_\text{B} > -6.5$ and $M_\text{V} > -5.5$. Recently, \citet{Johnson+2017} derived much stricter limits of $M_\text{U} > -3.8$, $M_\text{B} > -3.1$,
	and $M_\text{V} > -3.8$ for the SN 2012fh, which was also classified as type Ic but with some uncertainty due to its rather late discovery. In Table\,\ref{tab:magexamples}, we
	have compiled the absolute broadband magnitudes for three WC and WO stars which have rather typical parameters with regard to their spectral subtype. To allow an easy comparison,
	their magnitudes have been scaled such as if all stars would have the same $\log L/L_\odot = 5.5$. Assuming this luminosity, a WC6 progenitor would fit within the derived limits 
	for SN 2004gt, while a similar progenitor for SN 2012fh would need to be as faint as $\log L/L_\odot \la 4.7$. However, the first column of Table\,\ref{tab:magexamples} shows that absolute
	magnitudes decrease significantly when considering WO stars. Just considering the photometry, both SN 2004gt and SN 2012fh could easily have a WO progenitor, but not any
	kind of WC progenitor unless their luminosity was really low. The complex detectability of SNe Ibc progenitors and the result that progenitors with higher masses must not be
	optically brighter than those lower masses was already noted by \citet{Yoon+2012} when comparing evolutionary tracks with the WC results from \citet{Sander+2012}.
	
	Given the fact that especially type Ic supernova progenitors need to be depleted not only in hydrogen, but also in helium, WO stars are the only proper candidates with
	regard to chemical composition. This is also supported by theoretical studies from \citet{Groh+2013}, who calculated synthetic spectra for a set of stellar evolution  
	models and deduced that - assuming there is no silent collapse - all type Ic SNe in their calculations stem from WO stars while most type Ib SNe stem from late-type WN stars
	with only a smaller WO contribution. While WN stars will be discussed in a subsequent paper, we can already conclude that the current observational constraints of type Ibc
	SNe and the theoretical considerations point to a picture where WC stars are not SN progenitors, but only WO stars are. This would be qualitatively in line with
	the calculations from \citet{Groh+2013}, but it remains to be seen whether the remaining amount of approximately $X_\text{He} = 0.25$ could be ``hidden'' in type Ic SNe, which
	is a matter of debate \citep[e.g.][]{Hachinger+2012,Dessart+2012,PM2014,Prentice+2018}. 
	\citet{Frey+2013} demonstrated that enhanced convection could severely deplete helium, reducing it down to $X_\text{He} \approx 0.15$ for their $27\,M_\odot$ helium star model,
	not even accounting for WR mass-loss. With $X_\text{He} = 0.15$ derived by \citet{Tramper+2015}, the WO2 star WR 102 seems to be the most promising progenitor candidate for a type Ic SN
	so far.	
	
	Backed up by the identification of RSGs with $M_\text{i} \la 18\,M_\odot$ as type II supernova progenitors \citep[e.g.][]{Smartt2009,Smartt2015}, the low-luminosity
	part of the WC sample is hard to explain via classical single-star evolution. In theory there is the concept of chemically
	homogeneous evolution \citep[e.g.][]{Maeder1987,Langer1992,HL2000}, but this should be less efficient at higher metallicities due to the stronger stellar winds leading to 
	mass and angular momentum loss, eventually spinning down the stars and thus reducing the mixing \citep[e.g.][]{YL2005,Brott+2011,Cui+2018}.
	
  For the remaining regime between $M_\text{i} \ga 18\,M_\odot$ and the upper mass limit of RSGs, a post-RSG scenario would be the classical
	choice, but potentially enhanced by a low-luminosity LBV stage during the path of the star back to the blue part of the HRD, that is
		\begin{equation}
	  \text{O} \rightarrow \text{RSG} \rightarrow \text{WNL} \left[  \leftrightarrow \text{LBV} \right] \rightarrow \text{WNE} \rightarrow \text{WC} \rightarrow \text{WO} \rightarrow \text{CC}\text{.}
	\end{equation}
	The lower luminosities for the bona fide LBVs and LBV candidates obtained with the distances inferred from Gaia DR2 parallaxes 
	by \citet{Smith+2018} would support such an additional LBV stage, but their isolation argued by the same authors would 
	immediately contradict this scenario. Moreover, the luminosities derived by \citet{Smith+2018} are based on variety of different original
	luminosity determinations. As we see, e.g., from the conflicting results for the WO2 stars (cf. Sect.\,\ref{subsec:notable}), 
	a robust luminosity determination requires not only an accurate distance determination, but also a sufficient current-generation stellar 
	atmosphere analysis. Such research has so far only been performed for a small subsample of LBVs and LBV candidates.
	
 	Hydrogen-free WR stars are also the progenitor candidates for long-duration Gamma Ray Bursts (L-GRBs), which are interpreted as
	so-called ``failed SNe'' \citep[][]{Woosley1993,MW1999}, in contrast to shorter GRBs which are attributed to white dwarf 
	and compact object mergers as recently confirmed by the simultaneous detection of a gravitational wave and a GRB event \citep{Abbott+2017LigoGRB}.
	The term ``failed'' might be misleading as it does not necessarily mean that there is no SN, but that there is not enough neutrino energy for
	driving the explosion. However, as \citet{BW1983} argue, an explosion driven by other mechanisms (e.g., rotation, nuclear burning) can still 
	occur at a slightly later stage. In fact there is strong observational evidence \citep[e.g.,][]{Hjorth+2003,WB2006} for connections between type Ic SNe
	with broad lines -- also termed ``hypernovae'' due to their high inferred kinetic energies -- and L-GRBs. 
	
	For appearing as a type Ic SN, the progenitor star must not have any hydrogen and also a significant helium depletion, bringing WC and especially
	WO stars into the focus. The present nondetection of SN Ibc progenitors in all cases with available pre-explosion images makes it unlikely that
	WR stars are the only progenitors \citep[see, e.g., the discussion in Sect.\,5.1 in][]{Smartt2009}. Less luminous and thus less massive ``stripped'' progenitors
	which may not appear as WR stars might be needed as well. Nonetheless, WC and WO stars should still feed this SN channel, if they explode at all. Moreover, apart
	from classical sub\-dwarfs, the massive WR stars are the only well-studied objects in the hydrogen-free or nowadays more often termed ``stripped-envelope'' regime, 
  while the inferred parameters for a lower mass regime \citep[e.g.,][]{Vink2017,Goetberg+2018} are highly uncertain, due to the lack of constraints for various 
	ingredients in the underlying models, and can so far only be benchmarked by one object assumed to belong to these kind of stars, namely the 
	so-called qWR (``quasi Wolf-Rayet'') star HD\,45166 \citep{Groh+2008}. With sufficient mass loss, such stripped stars ranging from $M_\text{i} \la 18\,M_\odot$ would 
	appear as WN stars and might evolve further to the WC stage. In our WC and WO sample, there is no clear evidence that any of the sample stars is the
	product of such a process, but especially for the WC9 objects with the lowest luminosities - in particular WR 119 - this could be an evolutionary
  channel that would have to explored in more detail.
		
	Essentially for the same reasons as in the discussion about quasi-homogenous evolution, GRBs are found and expected to form at low metallicity \citep[e.g.,][]{YL2005,WH2006,Yoon+2006}, 
	explaining why they are often found at noticeable redshift. Nonetheless, GRB candidates in the form of fast-rotation WR stars have also been proposed and
	searched for at solar and near-solar metallicity \citep{Petrovic+2005,Vink+2011,Graefener+2012GRB}.	In \citet{Sander+2012}, we discussed the two WO2 stars as potential
	GRB candidates due to the high rotational velocity adopted for reproducing their observed spectra. Later, \citet{Tramper+2015} demonstrated that such a significant rotational
	velocity ($\sim\!1000\,$km/s) is not needed to reproduce the observed broad line profiles for the Galactic WO2 stars. However, \citet{Shenar+2014} showed that unless the wind significantly co-rotates,
	the emission line spectrum of a WR star with a dense wind is hardly affected, even if the star itself rotates with velocities up to $5000\,$km/s. Although unlikely as it
	would require to invent a mechanism that has lead to spun-up of the star, this means that Galactic WR stars essentially would be able to ``hide'' their rotation to a normal
	spectral analysis, and methods like polarization measurements are required. As long as this has not been done, the possibility of an L-GRB during CC for Galactic WO stars like
	WR 102 and WR 142 might not be likely, but cannot be ruled out either.

\section{Summary and conclusions}
  \label{sec:conclusions}
  
	Building on the parallaxes from Gaia DR2 and the distances inferred by \citet{Bailer-Jones+2018}, we have revised the stellar
	and wind parameters of the single Galactic WC and WO stars. Focussing on the impact of the new distances, only parameters depending on the 
	distances were updated compared to the analysis of \citet{Sander+2012}. The new results
	revealed that the assumption of roughly same $M_\varv$ per WC subtype is not justified, perhaps with the exception of the WC4 subtype.
	The luminosities of the Galactic WC stars are found to span a range from approximately $\log L/L_\odot = 4.9$ to $6.0$ with one outlier (WR 119)
	having $\log L/L_\odot = 4.7$. The most luminous star in our sample is the WC/WN transition-type star WR 126 with $\log L/L_\odot \approx 6.1$,
	which might not be a single object.
	
	For comparison, we also performed the same calculations for the three Galactic WO stars analyzed by \citet{Tramper+2015}.
  The parameters of those WO2 stars, which could be obtained from both sources, turned out to differ significantly in luminosity and
	all parameters depending on $L$. Since both studies used rather different atmosphere models, only a complete revision of the spectral
	analysis with new atmosphere models that resolve the issues of the previously applied ones can help to resolve this discrepancy.

	As a consequence of the new luminosities, also the mass-loss rates of the Galactic WC stars had to be revised. Adopting a (micro-)clumping
	factor of $D = 10$ for the whole sample, they cover the range between $\log \dot{M} = -5.1$ and $-4.1$, with a linear regression yielding $\dot{M} \propto L^{0.68}$. The two WO2 stars
	have mass-loss rates that are about $0.5$\,dex or $0.3$\,dex lower compared to WC stars with the same luminosity, depending on whether one
	uses the analyses from \citet{Sander+2012} or \citet{Tramper+2015}. We also found an essentially linear relation between 
	the modified wind momentum $D_\text{mom}$ and the revised WC luminosities, that is $D_\text{mom} \propto L$. Interpreting this with the mCAK theory
	would yield $\alpha_\text{eff} = 1$, while \citet{GH2005} numerically obtained $\alpha_\text{eff} \approx 0$. We take this as a further confirmation
	that current CAK-like concepts essentially fail do describe Wolf-Rayet winds. Obviously, a further investigation about the driving of WR winds 
	is needed.
	
\begin{table}
  \caption{Mean WC/WO star parameters per subtype}
  \label{tab:wcmean}	
  \centering

  \begin{tabular}{crccccc}
      \hline
      \hline
\multicolumn{1}{c}{Subtype\!\!\!\!} & \multicolumn{1}{c}{$T_{*}$\rule[0mm]{0mm}{3.5mm}} & \multicolumn{1}{c}{$\varv_{\infty}$} & \multicolumn{1}{c}{$R_{*}$} & \multicolumn{1}{c}{$\log\,\dot{M}$\tablefootmark{a} } & \multicolumn{1}{c}{$\log\,L$} & \multicolumn{1}{c}{$M_{\mathrm{WC}}$\tablefootmark{b}} \\
& \multicolumn{1}{c}{[kK]} & \multicolumn{1}{c}{[km/s]} & \multicolumn{1}{c}{[$R_{\odot}$]} & \multicolumn{1}{c}{[$M_{\odot}$/yr]} & \multicolumn{1}{c}{[$L_{\odot}$]} & \multicolumn{1}{c}{[$M_{\odot}$]} \\
      \hline
WO2 \rule[0mm]{0mm}{3mm} \!\!\!\! & $200$ & $5000$ & $0.7$ & $-4.96$ & $5.8$ & $22$ \\
WC4 \rule[0mm]{0mm}{3mm} \!\!\!\! & $117$ & $3310$ & $0.9$ & $-4.67$ & $5.2$ & $10$ \\
WC5 \rule[0mm]{0mm}{3mm} \!\!\!\! & $83$ & $2780$ & $3.2$ & $-4.39$ & $5.6$ & $18$ \\
WC6 \rule[0mm]{0mm}{3mm} \!\!\!\! & $78$ & $2270$ & $3.6$ & $-4.47$ & $5.7$ & $18$ \\
WC7 \rule[0mm]{0mm}{3mm} \!\!\!\! & $71$ & $2010$ & $4.0$ & $-4.57$ & $5.6$ & $17$ \\
WC8 \rule[0mm]{0mm}{3mm} \!\!\!\! & $60$ & $1810$ & $6.3$ & $-4.53$ & $5.6$ & $18$ \\
WC9 \rule[0mm]{0mm}{3mm} \!\!\!\! & $44$ & $1390$ & $8.7$ & $-4.66$ & $5.4$ & $13$ \\[0.5mm]
     \hline
  \end{tabular}
  \tablefoot{
    \tablefoottext{a}{All mass-loss rates are based on models from \citet{Sander+2012} which are calculated with a clumping factor of $D = 10$}
    \tablefoottext{b}{The masses of the WC stars are calculated from luminosities using the relation by \citet{Langer89mass}}
  }
\end{table}
	
	Mean WC parameters per subtype are listed in Table\,\ref{tab:wcmean}. We see that earlier WC subtypes haves higher $T_\ast$ and $\varv_\infty$.  
	A general trend of increasing radii with later subtypes can be noticed. Regarding luminosities and masses, only the WC4 stars show considerably lower values
	while there is no clear trend between the later WC subtypes with the WC9 slightly falling out. This is due to the result that only a smaller
	part of WC9 stars has a higher luminosity, while the majority has $\log L/L_\odot < 5.3$.

\begin{table}
  \caption{Suggested single-star evolution scenarios based revised WC/WO parameters. The mass ranges should be taken as rough estimates inferred from
	         comparing empirical luminosities with current evolutionary calculations instead of hard limits.}
  \label{tab:wcevol}
  \centering
	\small
  \begin{tabular}{r c l}
    \hline\hline
    \multicolumn{2}{c}{$M_{\mathrm{i}}$ [$M_{\odot}$] \rule[0mm]{0mm}{3mm}}  
                             &  \multicolumn{1}{c}{evolutionary path} \\ \hline
    \rule[0mm]{0mm}{3mm}     
      8~--~18	&	&		OB $\rightarrow$	RSG	 [$\rightarrow$	BSG	] $\rightarrow$	SN\,II		\\
     18~--~35	&	&		 O $\rightarrow$	RSG	$\rightarrow$		WNL  	[ $\leftrightarrow$ LBV ]	$\rightarrow$	WN  $\rightarrow$	WC \\
		          & &    \hspace{15.6em}          $\rightarrow$	WO   $\rightarrow$ CC	\\
     35~--~80	&	& 	 O $\rightarrow$	Of/WNH	$\leftrightarrow$	LBV	 $\rightarrow$	WN  $\rightarrow$	WC $\rightarrow$	WO $\rightarrow$ CC	\\
       $>$~80	&	&	   O $\rightarrow$	Of/WNH	$\leftrightarrow$	LBV	[$\rightarrow$  WNH] $\rightarrow$ SN\,IIn 	\\
    \hline
  \end{tabular}
\end{table}	

  In this work we did not update any of the underlying atmosphere analyzes performed for the sample stars. Thus, for an in-depth look at a
	particular object, we recommend to use the latest generation of atmospheres models, which have undergone various improvements since
	\citet{Sander+2012}. With regards to PoWR, an updated grids of atmosphere models with ready-to-use synthetic spectra for WC stars at Galactic 
	and LMC (not available in 2012) metallicity have been published on the PoWR website\footnote{\texttt{www.astro.physik.uni-potsdam.de/PoWR}}.
	
	Moreover, the situation in terms of radius and temperature discrepancies for earlier WC subtypes might be improved when
  a more accurate velocity and thus density stratification is assumed, which would require further updates of the atmosphere model code. 
  Recently, \citet{Grassitelli+2018} showed that it is possible to connect a stellar structure model smoothly to the hydrodynamically 
 consistent model for a WC5 star from \citet{GH2005}. While the calculation of such complex and time-consuming
 models is not an option when analyzing a whole population, a subset of representative models could
 provide valuable constraints such as velocity fields tailored to different WC subtypes.
	
  The current evolutionary status of the Galactic WC and WO stars as evolved, core-helium burning objects is solid, but the path toward this
	status remains uncertain. Our revised luminosities set new constraints, and, together with our recent results, slightly change the picture compared
	to \citet{Sander+2012}. For $\log L/L_\odot > 5.4$, current single star evolutionary calculations predict the formation of WC and WO stars at $Z_\odot$,
	although failing to reproduce the temperatures for the later subtypes, which might be due to the occurrence of inflated envelopes 
	\citep[e.g.,][]{Graefener+2012,Graefener+2017}. We conclude that while some of the stars in our sample
	might not have formed via the single star channel, the loss of mass without a binary companion is a viable channel to form WC and WO stars at $Z_\odot$.
	Suggested scenarios are summarized in Table\,\ref{tab:wcevol} and discussed in more detail in Sect.\,\ref{sec:evol}.
	Binary evolution is required to explain the lower luminosity WC stars and could also be responsible for a fraction of the higher luminosity stars. 
	The revision of wind mass-loss rates and better knowledge about
	episodic mass loss might still change this picture in the future, and could perhaps reopen the single star channel in the lower luminosity regime.

	A particular challenging object is the WC9d star WR 119. With a luminosity of only $\log L/L_\odot = 4.7$ and a current mass of approximately $6\,M_\odot$, 
	the star is likely a product of binary evolution, although there is no clear evidence for a close binary system. The system might be an interesting
	test case as it could be the evolved status of a so-called ``stripped envolope'' star, resulting from a primary with $M_\text{i} < 18\,M_\odot$ that 
	lost its hydrogen envelope to a companion and eventually from a WN to a WC stage.
	
	With masses between approximately $8\,$ and $30\,M_\odot$ for the remaining sample, the Galactic WC and WO stars are progenitors of massive black holes. The observed upper 
	limit of $\log L/L_\odot \approx 5.1$ for type II SNe progenitors and only one possible detection of a type Ibc progenitor opens the possibility
	that most massive stars, including our sample stars, collapse silently at the end of their lifetime, that is without a visible SN outburst. 
	However, if all WC stars evolve into a WO stage eventually, their absolute magnitudes in the U, B, and V filters 
	would be low enough to have escaped current progenitor detections. Thus, we conclude that lack of type Ibc SN progenitors does not
	necessarily mean that WR stars do collapse silently, although this scenario is evenly possible. Furthermore, the only class of classical 
	Wolf-Rayet stars that would be viable progenitor candidates for type Ic SNe, are the WO stars.

\begin{acknowledgements}
	The authors would like to thank the referee, C.~Georgy, for fruitful comments and suggestions that helped
	to increase the quality of this paper. A.A.C.S. would further like to acknowledge helpful discussions and suggestions
	with P.M.~Williams and J.S.~Vink.
  This research has made use of the SIMBAD database and the VizieR 
  catalog access tool, both operated at the CDS, Strasbourg, France. 
	This work has made use of data from the European Space Agency (ESA) mission \textit{Gaia}
	(\url{https://www.cosmos.esa.int/gaia}), processed by the \textit{Gaia} Data Processing 
	and Analysis Consortium (DPAC,\url{https://www.cosmos.esa.int/web/gaia/dpac/consortium}). 
	Funding for the DPAC has been provided by national institutions, in particular the institutions
  participating in the \textit{Gaia} Multilateral Agreement. The first author 
  of this work (A.A.C.S.) is supported by the Deutsche Forschungsgemeinschaft (DFG) under grant HA 1455/26 and would like
	to thank STFC for funding under grant number ST/R000565/1.
	T.S.\ acknowledges funding from the European Research Council (ERC) under the European Union's DLV\_772225\_MULTIPLES Horizon 2020
	research and innovation program. V.R.\ is grateful for financial support from 
	the Deutsche Akademische Austauschdienst (DAAD) as part of the Graduate School Scholarship Program. L.M.O. acknowledges support from the 
	DLR under grant 50 OR 1809.
	
\end{acknowledgements}


\bibliographystyle{aa} 
\bibliography{galwcgaia}

\end{document}